\documentclass{jnmp}

\usepackage{amsmath}

\JNMPnumberwithin{equation}{section}

\newtheorem{proposition}{Proposition}

\begin{document}

\renewcommand{\evenhead}{P~Bracken and A~M~Grundland}
\renewcommand{\oddhead}{Solutions of the Generalized Weierstrass Representation}

\thispagestyle{empty}

\FirstPageHead{9}{3}{2002}{\pageref{bracken-firstpage}--\pageref{bracken-lastpage}}{Article}

\copyrightnote{2002}{P~Bracken and A~M~Grundland}

\Name{Solutions of the Generalized Weierstrass\\ Representation
in Four-Dimensional\\ Euclidean Space}
\label{bracken-firstpage}

\Author{P~BRACKEN~$^{\star\dag}$ and
A~M~GRUNDLAND~$^{\star\ddag}$}

\Address{$^\star$~Centre de Recherches Math\'{e}matiques, Universit\'{e} de Montr\'{e}al, \\
~~2920 Chemin de la Tour,  Pavillon Andr\'{e} Aisenstadt,  C. P. 6128 Succ. Centre Ville,   \\
~~Montr\'{e}al, QC, H3C 3J7 Canada   \\
~~E-mail: bracken@mathstat.concordia.ca, \ grundlan@CRM.UMontreal.ca\\[10pt]
$^\dag$~Department of Mathematics, Concordia University,\\
~~Montr\'eal, QC, H4B 1R6, Canada\\[10pt]
$^{\ddag}$~Department of Mathematics,
Universit\'e du Qu\'ebec, \\
~~Trois-Rivieres, QC, G9A 5H7, Canada}

\Date{Received January 5, 2002; Revised  February 2, 2002;
Accepted April 15, 2002}

\begin{abstract}
\noindent
Several classes of solutions of
the generalized Weierstrass system, which
induces constant mean curvature
surfaces into four-dimensional Euclidean space are constructed.
A gauge transformation allows us to simplify the
system considered and derive factorized classes of solutions.
A reduction of the generalized Weierstrass system to decoupled $CP^1$ sigma
models is also considered.
A new procedure for
constructing certain classes of solutions, including elementary solutions
(kinks and bumps) and multisoliton solutions
is described in detail. The constant mean curvature
surfaces associated with different
types of solutions are presented.
Some physical interpretations of the results obtained
in the area of string theory are given.
\end{abstract}

\section{Introduction}

It has been shown~[1] that Weierstrass representations are very
useful and suitable tools for the systematic study of minimal
surfaces immersed in $n$-dimensional spaces. This subject has a
long and rich history. It has been extensively investigated since
the initial works of Weierstrass~[2] and Enneper~[3] in the middle
of the nineteenth century on systems inducing minimal surfaces in
${\mathbb R}^3$. In the literature there exists a great number of
applications of the Weierstrass representation to various domains
of Mathematics, Physics, Chemistry and Biology. In particular in
such areas as quantum field theory~[4], statistical physics~[5],
chemical physics, fluid dynamics and membranes~[6], minimal
surfaces play an essential role. More recently it is worth
mentioning that works by Kenmotsu~[7], Hoffmann~[8], Osserman~[9],
Budinich~[10], Konopelchenko~[11,~12] and Bobenko~[13,~14] have
made very significant contributions to constructing minimal
surfaces in a systematic way and to understanding their intrinsic
geometric properties as well as their integrable dynamics. The
type of extension of the Weierstrass representation which has been
useful in three-dimensional applications to multidimensional
spaces will continue to generate many additional applications to
physics and mathematics. According to~[15] integrable deformations
of surfaces are generated by the Davey--Stewartson hierarchy of
$2+1$ dimensional soliton equations. These deformations of
surfaces inherit all the remarkable properties of soliton
equations. Geometrically such deformations are characterised by
the invariance of an infinite set of functionals over surfaces,
the simplest being the Willmore functional.

In recent years, major developments in the area of the low
dimensional sigma model have been shown~[16] to be of use in
generating two-dimensional surfaces immersed in
multidimensional-space. There are links between this model and
other models, such as the non Abelian Chern--Simons theories which
have been of interest recently in condensed matter physics~[17].
In fact the Chern--Simons gauged Landau--Ginsburg model plays the
essential role of an effective theory for the Fractional Quantum
Hall Effect. There exists a link between non Abelian Chern--Simons
theories and the nonlinear sigma model, which is related to
minimal surfaces. For example, in~[18], a simple method was
proposed to obtain completely integrable systems in
$(2+1)$-dimensions from classes of non Abelian Chern--Simons field
theory. In this sense completely integrable systems are seen as
particular gauge choices in which the theory is formulated.
Moreover linear spectral problems are  naturally related to the
geometrical constraints imposed on the target space. Among several
possibilities for building up integrable deformations of
$(2+1)$-dimensional surfaces, multidimensional integrable spin
field systems can be used to realize integrable deformations of
surfaces. A more general $(2+1)$-dimensional integrable spin model
is described by the pair of equations~[17],
\begin{gather*}
\vec{S}_{t} + \vec{S} \wedge\left \{ (b+1) \vec{S}_{ss} - b \vec{S} \right\}
+ b u_{t} \vec{S}_{t} + (b+1) u_{s} \vec{S}_{s} =0,\\
u_{st} = \vec{S} \cdot ( \vec{S}_{s} \wedge \vec{S}_{t}),
\end{gather*}
where $s$ and $t$ are real or complex variables, $b$ a real
constant, $\vec{S} = (S_{1}, S_{2}, S_{3})$ is the spin
field vector, $\vec{S}^{2} =1$ and $u$ is a scalar function. These
represent one of the $(2+1)$-dimensional integrable generalizations of the
isotropic Landau--Lifshitz equation
\[
\vec{S}_{t} = \vec{S} \wedge \vec{S}_{xx}.
\]

The process of unifying gravity, supersymmetry and gauge theories
leads to supergravity theories such that the number of
supersymmetries goes from one to eight. Unfortunately quantum
Einstein gravity is non renormalizable~[4,~5]. A quantum theory of
gravity should therefore be a nonlocal quantum field theory.
Modern superstring theory draws together many concepts of field
theory, for example gauge symmetry, supersymmetry, effective
actions and the nonlinear sigma models. An extension of the
Polyakov string integral over multidimensional spaces would be of
great interest~[15]. In fact the string action in a non trivial
gravitational background takes the form of a non linear sigma
model or a~generalization of it. Conformal invariance plays a
fundamental role in perturbative string theory and results in deep
connections between strings and the nonlinear sigma model. Of
particular physical interest is the geometrical nature of the
interaction in the nonlinear sigma model which has consequences
such as the geometrical nature of the counterterms which are
required for renormalizability, the existence of topologically non
trivial field configurations such as solitons and gauge invariance
of four-dimensional quantum field theories.

In this paper we construct several classes of solutions
of the Weierstrass representation inducing
constant mean curvature (CMC) surfaces immersed in
Euclidean four-dimensional space.
This paper is an extension of previous papers~[19,~20] which
concern the Weierstrass representation for
CMC-surfaces immersed in Euclidean three-dimensional space.
This representation has recently been introduced by
Konopelchenko and Landolfi~[1] and will be referred
to as the KL system. Their formulas are the starting
point of our analysis, namely, they consider a first order
nonlinear system of two-dimensional Dirac type equations for
four complex valued functions
$\psi_{\alpha}$ and $\varphi_{\alpha}$ $(\alpha=1,2)$.
This system can be written as follows
\begin{gather}
\partial \psi_{\alpha} = p \varphi_{\alpha},
\qquad
\bar{\partial} \varphi_{\alpha} = - p \psi_{\alpha},  \qquad \alpha=1,2, \nonumber\\
p = \sqrt{u_{1} u_{2}} ,
\qquad
u_{\alpha} = |\psi_{\alpha}|^{2} + |\varphi_{\alpha}|^{2}
\end{gather}
and their complex conjugate equations. We denote the derivatives
in abbreviated form by $\partial = \partial/ \partial z$,
$\bar{\partial} = \partial / \partial \bar{z}$ and the bar denotes
complex conjugation. Note that eight of sixteen first
derivatives of the fields $\psi_{\alpha}$ and $\varphi_{\alpha}$
appearing in the KL system (1.1) are given in terms of the
complex functions $\psi_{\alpha}$ and $\varphi_{\alpha}$ only.
These functions $\psi_{\alpha}$ and
$\varphi_{\alpha}$ are invariant
under the multiplication factor minus one.
The system (1.1) possesses several
conservation laws such as
\begin{gather}
\partial (\psi_{\alpha} \psi_{\beta}) + \bar{\partial}
(\varphi_{\alpha} \varphi_{\beta})=0,
\qquad
\partial ( \psi_{\alpha} \bar{\varphi}_{\beta}) -
\bar{\partial} (\varphi_{\alpha} \bar{\psi}_{\beta})=0,
\qquad \alpha \neq \beta=1,2.
\end{gather}
As a consequence of these conserved quantities there exist four
real valued functions $X_{i} (z, \bar{z})$, $i=1, \ldots, 4$,
which can be interpreted as coordinates for a surface immersed in
Euclidean space ${\mathbb R}^{4}$. The coordinates of the position
vector ${\bf X}= \left(X^{1}, X^{2}, X^{3}, X^{4}\right)$ of
a~CMC-surface in ${\mathbb R}^4$ are given by~[1]
\begin{gather}
X^{1} = \frac{i}{2} \int_{\gamma} \left[ ( \bar{\psi}_{1}
\bar{\psi}_{2} + \varphi_{1} \varphi_{2}) \, dz'
- (\psi_{1} \psi_{2} + \bar{\varphi}_{1} \bar{\varphi}_{2} ) \, d \bar{z}'\right],\nonumber\\
X^{2} = \frac{1}{2} \int_{\gamma} \left[ (\bar{\psi}_{1}
\bar{\psi}_{2} - \varphi_{1} \varphi_{2}) \, dz'
+ (\psi_{1} \psi_{2} - \bar{\varphi}_{1} \bar{\varphi}_{2}) \, d \bar{z}'\right], \nonumber\\
X^{3} = - \frac{1}{2} \int_{\gamma} \left[ (\bar{\psi}_{1}
\varphi_{2} + \bar{\psi}_{2} \varphi_{1}) \, dz'
+ ( \psi_{1} \bar{\varphi}_{2} + \psi_{2} \bar{\varphi}_{1}) \, d \bar{z}'\right], \nonumber\\
X^{4} = \frac{i}{2}  \int_{\gamma} \left[ ( \bar{\psi}_{1}
\varphi_{2} - \bar{\psi}_{2} \varphi_{1}) \, dz' - (\psi_{1}
\bar{\varphi}_{2} - \psi_{2} \bar{\varphi}_{1}) \, d
\bar{z}'\right],
\end{gather}
where $\gamma$ is any closed contour in the complex plane $\mathbb C$. Due to the
conservation laws (1.2)
the integrals appearing in system (1.3) do not depend upon the trajectory
of contour $\gamma$ in $\mathbb C$, but depend upon the endpoints.
The differentials of equations
(1.3) are exact ones. The mean and the Gaussian
curvatures and the first and second fundamental forms of
the surface immersed in ${\mathbb R}^{4}$ are given by~[1]
\begin{gather}
{\bf H}^2 = 4  \frac{|p|^2}{u_{1} u_{2}},
\qquad
K= - p^{-2} \partial \bar{\partial} \ln p,   \nonumber\\
ds^{2} = u_{1} u_{2} \, dz \, d \bar{z},
\qquad II= (\partial^{2} {\bf r} | {\bf n}) dz^{2}
+ (\partial \bar{\partial} {\bf r} | {\bf \bar{n}}) dz \, d \bar{z}
+ ( \bar{\partial}^{2} {\bf r} | {\bf n}) d \bar{z}^{2}
\end{gather}
in conformal coordinates, respectively. Here the vector ${\bf n}$
is the unit normal vector to a~surface which satisfies $(\partial
{\bf r} | {\bf n})=0$, $(\bar{\partial} {\bf r}| {\bf n})=0$,
${\bf n}^{2} =1$ and the bracket $(\;|\;)$ denotes the standard
scalar product in ${\mathbb R}^{4}$.

In our investigations it is more convenient to introduce two
new dependent complex variables which link the Weierstrass
representation with a second order overdetermined system
of PDEs. This link allows us to establish several
useful transformations in order to simplify the structure
of the KL system (1.1) and to construct several classes of solutions.

We define two new complex variables
given by
\begin{gather}
\xi_{\alpha} = \frac{\psi_{\alpha}}{\bar{\varphi}_{\alpha}},  \qquad
\alpha= 1, 2.
\end{gather}
The equations (1.3) can be written
in equivalent form~[1] in terms of $\xi_{\alpha}$ as
\begin{gather}
{\bf X} = \int^{z} {\rm Re}\, (\vartheta {\bf G} \, dz),
\end{gather}
where the functions $\vartheta$ and $G( z, \bar{z})$ are given by
\begin{gather}
\vartheta^{2} =
-  \frac{4 \partial \xi_{1} \partial \xi_{2}}{{\bf H}^2 (1 + |\xi_{1}|^{2})^{2}(1+ |\xi_{2}|^{2})^{2}},
\nonumber\\
{\bf G } ( z, \bar{z}) =
[ (1 + \bar{\xi}_{1} \bar{\xi}_{2}), i ( 1 - \bar{\xi}_{1} \bar{\xi}_{2}),
i ( \bar{\xi}_{1} + \bar{\xi}_{2}), \bar{\xi}_{1} - \bar{\xi}_{2} ]
\end{gather}
and the complex functions $\xi_{1}$ and $\xi_{2}$ obey the
second order partial differential equation
\begin{gather}
-2 \partial (\ln |{\bf H}|) + \frac{\partial \bar{\partial} \bar{\xi}_{1}}{\bar{\partial} \bar{\xi}_{1}}
- \frac{2 \xi_{1} \partial \bar{\xi}_{1}}{1+ |\xi_{1}|^{2}}
+ \frac{\partial \bar{\partial} \bar{\xi}_{2}}{\bar{\partial} \bar{\xi}_{2}}
- \frac{2 \xi_{2} \partial \bar{\xi}_{2}}{1 + |\xi_{2}|^{2}} = 0.
\end{gather}

In particular, if $\psi_{2} = \epsilon \psi_{1}$
and $\varphi_{2} = \epsilon \varphi_{1}$ for
$\epsilon = \pm 1$ hold, then the KL system
(1.1) is reduced to the generalized Weierstrass (GW) system
inducing CMC-surfaces in $\mathbb R^{3}$
\begin{gather}
\partial \psi_{1} = p \varphi_{1}, \qquad
\bar{\partial} \varphi_{1} = -p \psi_{1}, \qquad
p = |\psi_{1}|^{2} + |\varphi_{1}|^2.
\end{gather}
Equation (1.8) then becomes the equation for the $CP^1$ sigma model
\begin{gather}
\partial \bar{\partial} \xi - \frac{2 \bar{\xi}}
{1 + |\xi|^{2}}  \partial \xi
\bar{\partial} \xi = 0
\end{gather}
and its conjugate, since $\xi_{1} = \xi_{2} = \xi$ by (1.5).
These limits characterize the properties of
solutions of the KL system (1.1).

In this paper we examine certain algebraic
and differential constraints of the first order,
compatible with the initial system of PDEs (1.1),
which allow us to simplify its structure.
In particular we focus upon constructing several
classes of multisoliton solutions of system (1.1)
which have not been found up to now.
In some cases new, interesting, CMC surfaces
are found in explicit form.

The paper is organized as follows.
In Section~2 we perform a reduction of the original system~(1.1)
to a certain overdetermined system of PDEs.
This overdetermined system allows us to simplify the
structure of the KL system by performing a rational
transformation for the functions $\psi_{\alpha}$ and
$\varphi_{\alpha}$ in~(1.1).
In Section~3 we introduce a gauge transformation
for the KL system and
we discuss the possibility of factorization of the associated KL system.
Furthermore a new procedure for constructing solutions to
the KL system is proposed.
We formulate several useful propositions
for building certain classes of solutions of the KL system.
Section~4 deals with a reduction of the KL system to
decoupled $CP^1$ sigma models.
In Section~5 we apply these
propositions in order to construct several
explicit solutions and their superpositions.
Based on these propositions we are
able to generate new classes of multisoliton solutions
and find their associated CMC surfaces.
Section~6 contains a simple example related to classical string
configurations in $\mathbb R^4$ and possible future
developments.

\section{A system associated\\ with the
Konopelchenko--Landolfi system}

Now we introduce a new system associated with (1.1)
which allows the construction of seve\-ral classes of
solutions for the KL system including elementary
and multisoliton solutions, which are presented in Section~5.

\begin{proposition}
If the complex valued functions $\psi_{\alpha}$ and
$\varphi_{\alpha}$ are solutions of KL
system~(1.1), then the rational functions defined by (1.5)
are solutions of the following overdetermined system,
\begin{gather}
\frac{(\partial \xi_{1})(\bar{\partial} \bar{\xi}_{1})}
{(1+ |\xi_{1}|^{2})^{2}} -
\frac{(\partial \xi_{2})(\bar{\partial} \bar{\xi}_{2})}
{(1+ |\xi_{2}|^{2})^{2}} = 0,
\\
-\frac{\partial \bar{\partial} \partial \xi_{\alpha}}
{2 \partial \xi_{\alpha}} +
\frac{\bar{\partial} \partial \xi_{\alpha}}{2 (\partial \xi_{\alpha})^{2}}
\partial^{2} \xi_{\alpha}
- \frac{\partial \bar{\partial} \bar{\xi}_{\alpha}}
{1+ |\xi_{\alpha}|^{2}} \xi_{\alpha}
+ \frac{\xi_{\alpha}^{2}(\partial \bar{\xi}_{\alpha})
(\bar{\partial} \bar{\xi}_{\alpha})}
{(1 + |\xi_{\alpha}|^{2})^{2} }  + \frac{\bar{\partial} \bar{\partial} \partial \xi_{\alpha}}
{2 \bar{\partial} \bar{\xi}_{\alpha}}   \nonumber\\
\qquad {}
- \frac{\bar{\partial} \partial \bar{\xi}_{\alpha}}
{2 (\bar{\partial} \bar{\xi}_{\alpha})^{2}}
\bar{\partial}^{2} \bar{\xi}_{\alpha}+
\frac{\bar{\partial} \partial \xi_{\alpha}}{1 + |\xi_{\alpha}|^{2}}
\bar{\xi}_{\alpha} -
\frac{\bar{\xi}_{\alpha}^{2} (\bar{\partial} \xi_{\alpha})
(\partial \xi_{\alpha})}{(1+ |\xi_{\alpha}|^{2})^{2}} = 0,
\qquad \alpha=1,2.
\end{gather}
\end{proposition}

\begin{proof} In fact, from (1.1) and making use of transformation (1.5), we get
\begin{gather}
u_{\alpha} = |\psi_{\alpha}|^{2} + |\varphi_{\alpha}|^{2}
= |\varphi_{\alpha}|^{2} \left(1 + |\xi_{\alpha}|^{2}\right),
\qquad
\alpha= 1,2.
\end{gather}
By differentiation of equations (1.5) with respect to $\partial$
and using system (1.1), we obtain
\begin{gather}
\partial \xi_{\alpha} = p \, \frac{u_{\alpha}}{\bar{\varphi}_{\alpha}^{2}}
= p \frac{\varphi_{\alpha}}{\bar{\varphi}_{\alpha}} \left(1 + |\xi_{\alpha}|^{2}\right).
\end{gather}
Taking the ratio of (2.4) with its complex conjugate, we get
\begin{gather}
\frac{\partial \xi_{\alpha}}{\bar{\partial} \bar{\xi}_{\alpha}}
= \frac{\varphi_{\alpha}^{2}}{\bar{\varphi}_{\alpha}^{2}}.
\end{gather}
Multiplying equation (2.4) by its complex conjugate
and solving for $p^2$, we can express $p^2$ in terms of $\xi_{\alpha}$ as
\begin{gather}
(\partial \xi_{\alpha})(\bar{\partial} \bar{\xi}_{\alpha})
= p^{2} \left(1 + |\xi_{\alpha}|^{2}\right)^{2}.
\end{gather}
So we have
\begin{gather}
p^{2} = \frac{(\partial \xi_{\alpha})(\bar{\partial} \bar{\xi}_{\alpha})}
{\left(1 + |\xi_{\alpha}|^{2}\right)^{2}},
\qquad
\alpha=1,2.
\end{gather}
Equating equations (2.4) for $\alpha = 1,2$ we obtain expression (2.1).
Using equation (2.5) we can write equations (2.7) in equivalent form
\begin{gather}
(i) \quad
p^{2} = \frac{\varphi_{\alpha}^{2} ( \bar{\partial} \bar{\xi}_{\alpha})^{2}}
{\bar{\varphi}_{\alpha}^{2} (1+ |\xi_{\alpha}|^{2})^{2}},
\qquad  (ii)
\quad
p^{2} = \frac{\bar{\varphi}_{\alpha}^{2} ( \partial \xi_{\alpha})^{2}}
{\varphi_{\alpha}^{2} (1 + |\xi_{\alpha}|^{2})^{2}},
\qquad
\alpha=1,2.
\end{gather}
Elimination of $p^{2}$ from (2.8), for different values of
$\alpha=1,2$, leads to equation (2.1). Eliminating $p^{2}$ from
equations (2.7) with $\alpha=1$ and (2.8ii) with $\alpha=2$ we get
the following relation
\begin{gather}
\varphi_{2}^{2} \frac{(\partial \xi_{1})(\bar{\partial} \bar{\xi}_{1})}
{(1+ |\xi_{1}|^{2})^{2}} - \bar{\varphi}_{2}^{2} \frac{(\partial \xi_{2})^{2}}
{(1 + |\xi_{2}|^{2})^{2}} =0.
\end{gather}
Similarly, from equations (2.7) with $\alpha=2$ and equations
(2.8ii) with $\alpha=2$, we have
\begin{gather}
\varphi_{2}^{2} ( \bar{\partial} \bar{\xi}_{2}) - \bar{\varphi}_{2}^{2}
(\partial \xi_{2}) = 0.
\end{gather}
Equations (2.9) and (2.10) have a nontrivial solution for
functions $\varphi_{2}^{2}$
and $\bar{\varphi}_{2}^{2}$ if the determinant of their coefficients
vanishes. This condition turns out to be exactly the condition (2.1).
Elimination of $p^{2}$ from equations
(2.7) and (2.8i)
with $\alpha=1$ and, next,
elimination of $p^{2}$ from (2.7) with $\alpha=2$
and (2.8ii) with $\alpha=1$ leads to
\begin{gather}
\varphi_{1}^{2} (\bar{\partial} \bar{\xi}_{1}) - \bar{\varphi}_{1}^{2}
(\partial \xi_{1}) = 0,
\\
\varphi_{1}^{2} \frac{(\partial \xi_{2})(\bar{\partial} \bar{\xi}_{2})}
{( 1+ |\xi_{2}|^{2})^{2}} - \bar{\varphi}_{1}^{2}
\frac{(\partial \xi_{1})^{2}}{(1+ |\xi_{1}|^{2})^{2}} = 0.
\end{gather}
The condition for the existence of nontrivial solutions
for $\varphi_{1}^{2}$ and $\bar{\varphi}_{1}^{2}$
of equations (2.11) and (2.12) is reduced to condition (2.1).
This means that we can take into account only two
equations, say (2.10) and (2.11), from the systems of equations
(2.9)--(2.12) since the determinant of
their coefficients vanishes whenever (2.1) holds.
This implies that equations (2.9)--(2.12) are linearly dependent.
A general solution for the system (2.10) and (2.11)
has the form
\begin{gather}
\varphi_{\alpha} = a_{\alpha} (\partial \xi_{\alpha})^{1/2},
\qquad
\bar{\varphi}_{\alpha} = \bar{a}_{\alpha} (\bar{\partial}
\bar{\xi}_{\alpha})^{1/2}.
\end{gather}
For (2.13) to be consistent with (2.9)--(2.12)
the functions $a_{\alpha} = a_{\alpha} (z, \bar{z})$
are real-valued so that
$a_{\alpha} = \bar{a}_{\alpha}$.
Substituting (2.13) into (1.5) we obtain
\begin{gather}
\psi_{\alpha} = a_{\alpha} \xi_{\alpha} (\bar{\partial} \bar{\xi}_{\alpha})^{1/2},
\qquad
\bar{\psi}_{\alpha} = a_{\alpha} \bar{\xi}_{\alpha} (\partial \xi_{\alpha})^{1/2}.
\end{gather}
Differentiating (2.13) with respect to $\bar{\partial}$
and substituting into the KL system (1.1), we obtain
\begin{gather*}
(\bar{\partial} a_{\alpha}) (\partial \xi_{\alpha})^{1/2}
+ \frac{1}{2} a_{\alpha} (\partial \xi_{\alpha})^{-1/2}
\partial \bar{\partial} \xi_{\alpha}\\
\qquad {} =
-p \xi_{\alpha} a_{\alpha} (\bar{\partial} \bar{\xi}_{\alpha})^{1/2}
= - \frac{(\partial \xi_{\alpha})^{1/2} (\bar{\partial} \bar{\xi}_{\alpha})^{1/2}}{1+ |\xi_{\alpha}|^{2}} \xi_{\alpha} a_{\alpha} (\bar{\partial}
\bar{\xi}_{\alpha})^{1/2}.
\end{gather*}
This means that the functions $\xi_{\alpha}$ satisfy the
second order differential equation
\begin{gather}
\partial \bar{\partial} \xi_{\alpha} +2 (\bar{\partial} \ln a_{\alpha})
(\partial \xi_{\alpha}) = -2 \frac{(\partial \xi_{\alpha})
(\bar{\partial} \bar{\xi}_{\alpha})}{1 + |\xi_{\alpha}|^{2}} \xi_{\alpha}.
\end{gather}
Similarly differentiation of (2.14) with respect to $\partial$
and use of the KL system (1.1) yields the
complex conjugate of (2.15), namely,
\begin{gather*}
\partial \bar{\partial} \bar{\xi}_{\alpha}
+ 2 ( \partial \, \ln a_{\alpha})
(\bar{\partial} \bar{\xi}_{\alpha}) = -2
\frac{(\partial \xi_{\alpha}) (\bar{\partial} \bar{\xi}_{\alpha})}
{1 + |\xi_{\alpha}|^{2}} \bar{\xi}_{\alpha}.
\end{gather*}
Equations (2.15) and their conjugates can be solved for
the quantities $\bar{\partial}(\ln a_{\alpha})$
and $\partial (\ln a_{\alpha})$, respectively,
\begin{gather}
(i) \quad
\bar{\partial} \ln a_{\alpha} =
- \frac{\bar{\partial} \partial \xi_{\alpha}}{2 \partial \xi_{\alpha}}
- \frac{\bar{\partial} \bar{\xi}_{\alpha}}{1+ |\xi_{\alpha}|^{2}}
\xi_{\alpha},
\nonumber\\
(ii)
\quad
\partial \ln a_{\alpha} = - \frac{\bar{\partial} \partial \bar{\xi}_{\alpha}}
{2 \bar{\partial} \bar{\xi}_{\alpha}} - \frac{\partial \xi_{\alpha}}
{1 + |\xi_{\alpha}|^{2}} \bar{\xi}_{\alpha}.
\end{gather}
By differentiation of (2.16i) with respect to $\partial$ and
(2.16ii) with respect to $\bar{\partial}$, the compatibility
condition for these derivatives results in (2.2).
Thus, for a given
solution $\xi_{\alpha}$ of system (2.1) and (2.2), the function $a_{\alpha}$
is determined uniquely by equations (2.16).
By use of (2.15) in (1.8) the second derivatives
can be eliminated from (1.8) to obtain an additional
constraint on the functions $a_{\alpha}$,
\begin{gather}
\partial \ln a_{1} + \frac{1}{1 + |\xi_{1}|^{2}}
( \bar{\xi}_{1} \partial \xi_{1} + \xi_{1} \partial \bar{\xi}_{1})
+ \partial \ln a_{2} + \frac{1}{1 + | \xi_{2}|^{2}}
(\bar{\xi}_{2} \partial \xi_{2} + \xi_{2} \partial
\bar{\xi}_{2} ) = 0
\end{gather}
and its complex conjugate. By virtue of (1.8) the condition (2.17)
is a necessary condition for the mean curvature $H$ to be constant,
according to (1.4).
\end{proof}

Note that, if the real-valued functions $a_{\alpha}$ are
holomorphic functions, then $a_{\alpha}$ are constant and (2.15) reduces
to two decoupled $CP^1$ sigma model equations (1.10).
The converse Proposition is also true.

\begin{proposition}
Suppose that the complex valued functions $\xi_{\alpha}$
are solutions of the overdetermined system (2.1) and (2.2)
and that the real-valued functions $a_{\alpha}$ are solutions of equations (2.16)
and (2.17). Then the complex functions $\varphi_{\alpha}$ and $\psi_{\alpha}$ determined by
\begin{gather}
\varphi_{\alpha} = a_{\alpha} (\partial \xi_{\alpha})^{1/2},
\qquad
\psi_{\alpha} = a_{\alpha} \xi_{\alpha} (\bar{\partial} \bar{\xi}_{\alpha})^{1/2},
\qquad
\alpha=1,2,
\end{gather}
are solutions of the KL system (1.1).
\end{proposition}

\begin{proof} For a given solution of (2.1), (2.2)
we assume that the functions $a_{\alpha}$ are
consistent with the compatibility conditions (2.16) and (2.17).
Differentiating $\varphi_{\alpha}$ in (2.18)
with respect to $\bar{\partial}$ we obtain
\begin{gather}
\bar{\partial} \varphi_{\alpha} = (\bar{\partial} a_{\alpha})
(\partial \xi_{\alpha})^{1/2} + \frac{1}{2} a_{\alpha} (\partial
\xi_{\alpha})^{-1/2} \bar{\partial} \partial \xi_{\alpha}.
\end{gather}
Using (2.16i) we can eliminate
the first derivative of $a_{\alpha}$ in equation (2.19) to obtain
\begin{gather*}
\bar{\partial} \varphi_{\alpha} = \left( - \frac{\bar{\partial}
\partial \xi_{\alpha}}{2 ( \partial \xi_{\alpha})} a_{\alpha} -
\frac{\bar{\partial} \bar{\xi}_{\alpha}} {1 + |\xi_{\alpha}|^{2}}
\xi_{\alpha} a_{\alpha}\right) (\partial \xi_{\alpha})^{1/2} +
\frac{1}{2} a_{\alpha}
 \: \frac{\bar{\partial} \partial \xi_{\alpha}}{(\partial \xi_{\alpha})^{1/2}}.
\end{gather*}
Next, making use of (2.7) and (2.18), we get
\[
\bar{\partial} \varphi_{\alpha}
=-\frac{(\bar{\partial} \bar{\xi}_{\alpha})^{1/2} (\partial \xi_{\alpha})^{1/2}}
{1 + |\xi_{\alpha}|^{2}} a_{\alpha} \xi_{\alpha}
( \bar{\partial} \bar{\xi}_{\alpha})^{1/2} = - p \psi_{\alpha}.
\]
So the functions $\varphi_{\alpha}$ satisfy the first equation in (1.1).
Differentiating $\psi_{\alpha}$ with respect
to $\partial$ in (2.18), we obtain
\begin{gather}
\bar{\partial} \psi_{\alpha} = (\partial \xi_{\alpha}) a_{\alpha}
(\bar{\partial} \bar{\xi}_{\alpha})^{1/2} + \xi_{\alpha}
(\partial a_{\alpha}) (\bar{\partial} \bar{\xi}_{\alpha})^{1/2}
+ \frac{1}{2} \xi_{\alpha} a_{\alpha} (\bar{\partial}
\bar{\xi}_{\alpha})^{-1/2} \partial \bar{\partial} \bar{\xi}_{\alpha}.
\end{gather}
Similarly, using equations (2.16ii) and (2.7),
we can eliminate the first derivatives in equation (2.20) and (2.18).
We have
\begin{gather*}
\bar{\partial} \psi_{\alpha}
= (\partial \xi_{\alpha}) a_{\alpha} ( \bar{\partial} \bar{\xi}_{\alpha})^{1/2}
+ \xi_{\alpha} ( \bar{\partial} \bar{\xi}_{\alpha})^{1/2}
\left(- \frac{\bar{\partial} \partial \bar{\xi}_{\alpha}}
{2 (\bar{\partial} \bar{\xi}_{\alpha})} - \frac{\partial \xi_{\alpha}}
{1 + |\xi_{\alpha}|^{2}} \bar{\xi}_{\alpha}\right) a_{\alpha}\\
\qquad {}+ \frac{1}{2} \xi_{\alpha} a_{\alpha}
(\bar{\partial} \bar{\xi}_{\alpha})^{-1/2}
\partial \bar{\partial} \bar{\xi}_{\alpha}
= \frac{(\partial \xi_{\alpha})^{1/2} (\bar{\partial} \bar{\xi}_{\alpha})^{1/2}}
{1+ |\xi_{\alpha}|^{2}} a_{\alpha} ( \partial \xi_{\alpha})^{1/2}
= p \varphi_{\alpha}.
\end{gather*}
This completes the proof.
\end{proof}

\section{Gauge transformations and factorization\\ of the associated
Konopelchenko--Landolfi system}

Now, we discuss certain new classes of multisoliton solutions
of the KL system (1.1) which can be obtained directly by applying
the transformation (1.5).
Firstly, we demonstrate that the KL system (1.1) admits a gauge transformation.
We introduce a new set of complex functions,
$\kappa_{\alpha}, \tau_{\alpha} : {\mathbb C}
\rightarrow {\mathbb C}$, which are related to the complex functions
$\psi_{\alpha}$ and $\varphi_{\alpha}$ by
\begin{gather}
\psi_{\alpha} = f_{\alpha} (z, \bar{z}) \kappa_{\alpha},
\qquad
\bar{\varphi}_{\alpha} = f_{\alpha} (z, \bar{z}) \bar{\tau}_{\alpha}
\end{gather}
for any complex functions $f_{\alpha} : {\mathbb C} \rightarrow {\mathbb C}$.

From equations (1.5), since $\psi_{\alpha}$,
$\bar{\varphi}_{\alpha}$ appear as a ratio, it is evident that the
transformation (3.1) leaves the functions $\xi_{\alpha}$ invariant
\begin{gather}
\xi_{\alpha} = \frac{\kappa_{\alpha}}{\bar{\tau}_{\alpha}}.
\end{gather}
This means that there exists a freedom which resembles a type of gauge
freedom in the definition of the functions, $\xi_{\alpha}$,
since the numerator and denominator of (1.5) can be multiplied
by any complex function. The main point is that it is not required
that the set of functions $\kappa_{\alpha}$ and $\bar{\tau}
_{\alpha}$ satisfy the original system (1.1), but that the ratio
of $\kappa_{\alpha}$ over $\bar{\tau}_{\alpha}$
has to satisfy the system (2.1) and (2.2).

\begin{proposition} Suppose that for any real holomorphic
functions, $g_{\alpha}$,
the complex functions, $\kappa_{\alpha}$,
and, $\tau_{\alpha}$, are related to the complex functions,
$\psi_{\alpha}$, and, $\varphi_{\alpha}$, as follows
\begin{gather}
\kappa_{\alpha} = g_{\alpha} \psi_{\alpha},
\qquad
\tau_{\alpha} = {g}_{\alpha} \varphi_{\alpha},
\qquad
\partial g_{\alpha} = 0.
\end{gather}
Then the functions $\kappa_{\alpha}$ and $\tau_{\alpha}$ are solutions of
the KL system (1.1) and have the form given by (2.14) provided
that the functions $\xi_{\alpha}$
are solutions of (2.1) and (2.2) and the real valued functions
$a_{\alpha} (z, \bar{z})$ have to satisfy the conditions (2.16) and (2.17).
\end{proposition}

\begin{proof} Note, from the fact that $\xi_{\alpha}$ in (3.2)
is invariant under the gauge function $f_{\alpha}$, it follows that
the function $p$ given by (2.7) is invariant as well.

From (3.3) and using (2.14) we can write that
\begin{gather}
\kappa_{\alpha}= g_{\alpha} a_{\alpha} \xi_{\alpha}
(\bar{\partial} \bar{\xi}_{\alpha})^{1/2},
\qquad
\tau_{\alpha} = {g}_{\alpha} a_{\alpha}
(\partial \xi_{\alpha})^{1/2}.
\end{gather}
Differentiating $\kappa_{\alpha}$ with respect to $\partial$,
we obtain
\begin{gather}
\partial \kappa_{\alpha} = (\partial g_{\alpha}) a_{\alpha}
\xi_{\alpha} (\bar{\partial} \bar{\xi}_{\alpha})^{1/2}
+ g_{\alpha} (\partial a_{\alpha}) \xi_{\alpha}
(\bar{\partial} \bar{\xi}_{\alpha})^{1/2} \nonumber\\
\phantom{\partial \kappa_{\alpha} =}{}+ g_{\alpha} a_{\alpha}
(\partial \xi_{\alpha})(\bar{\partial} \bar{\xi}_{\alpha})^{1/2} +
\frac{1}{2}g_{\alpha} a_{\alpha} \xi_{\alpha} (\bar{\partial}
\bar{\xi}_{\alpha})^{-1/2}
\partial \bar{\partial} \bar{\xi}_{\alpha}
\end{gather}
Substitution of the derivative $\partial a_{\alpha}$
obtained from (2.16) into (3.5), leads to (3.5) becoming
\begin{gather}
\partial \kappa_{\alpha}
= (\partial g_{\alpha}) a_{\alpha} \xi_{\alpha}
(\bar{\partial} \bar{\xi}_{\alpha})^{1/2}
- g_{\alpha} a_{\alpha} \frac{(\partial \xi_{\alpha})
(\bar{\partial} \bar{\xi}_{\alpha})^{1/2}}{1 + |\xi_{\alpha}|^{2}}
|\xi_{\alpha}|^{2}
+ g_{\alpha} a_{\alpha} (\partial \xi_{\alpha})
(\bar{\partial} \bar{\xi}_{\alpha})^{1/2}
\nonumber\\
\phantom{\partial \kappa_{\alpha}} {}= (\partial g_{\alpha})
a_{\alpha} \xi_{\alpha} (\bar{\partial} \bar{\xi}_{\alpha})^{1/2}
+ \left(\frac{(\partial \xi_{\alpha})^{1/2} (\bar{\partial}
\bar{\xi}_{\alpha})^{1/2}} {1 + |\xi_{\alpha}|^{2}}\right)
\left(g_{\alpha} a_{\alpha} (\partial \xi_{\alpha})^{1/2}\right).
\end{gather}
Clearly, if $\partial g_{\alpha} = 0$, equation (3.6) takes the
form of the first equation of the KL system~(1.1), $\partial
\kappa_{\alpha}= p \tau_{\alpha}$, as required.

Differentiating $\tau_{\alpha}$ with respect to $\bar{\partial}$,
we obtain
\begin{gather*}
\bar{\partial} \tau_{\alpha}=
\bar{\partial} g_{\alpha} \, a_{\alpha} (\partial \xi_{\alpha})^{1/2} +
{g}_{\alpha} (\bar{\partial} a_{\alpha}) (\partial \xi_{\alpha})^{1/2}
+ \frac{1}{2} (\partial \xi_{\alpha})^{-1/2}(\bar{\partial} \partial
\xi_{\alpha}) g_{\alpha} a_{\alpha}
\\
\phantom{\bar{\partial} \tau_{\alpha}}{}= a_{\alpha}
(\bar{\partial} g_{\alpha}) (\partial \xi_{\alpha})^{1/2} -
g_{\alpha} a_{\alpha} \frac{(\partial \xi_{\alpha})^{1/2}
\bar{\partial} \bar{\xi}_{\alpha}}{1 + |\xi_{\alpha}|^{2}}
\xi_{\alpha}.
\end{gather*}
Since $g_{\alpha}$ are holomorphic functions, the above equations simplify to
\begin{gather}
\bar{\partial} \tau_{\alpha} =
- \frac{(\partial \xi_{\alpha})^{1/2} (\bar{\partial} \bar{\xi}_{\alpha})^{1/2}}
{1 + |\xi_{\alpha}|^{2}} \left(g_{\alpha} a_{\alpha} \xi_{\alpha}
\left(\bar{\partial} \bar{\xi}_{\alpha}\right)^{1/2}\right), \qquad \alpha=1,2.
\end{gather}
The first factor on the right hand
side of (3.7) is just $p$. Hence the
second factor of (3.7) is the expression given by (3.4)
for the function $\kappa_{\alpha}$. Thus
(3.7) is the second equation in the KL system,
$\bar{\partial} \tau_{\alpha}=-p \kappa_{\alpha}$,
which completes the proof.
\end{proof}

Now we discuss in detail certain classes of solutions to (1.1) that can be
obtained from the transformation (1.5) by subjecting the
systems (2.1) and (2.2) to the following algebraic constraints
\begin{gather}
|\xi_{\alpha}|^{2} = 1, \qquad \alpha = 1,2.
\end{gather}
This implies that the functions $\xi_{1}$ and $\xi_{2}$ can differ
only by a phase when these functions are represented in polar
coordinates in the complex plane $\mathbb C$. By virtue of (3.8)
equation~(2.15) becomes
\begin{gather}
\partial \bar{\partial} \xi_{\alpha} +  {\partial \xi_{\alpha}
\bar{\partial} \bar{\xi}_{\alpha}} \xi_{\alpha}
+  2 \bar{\partial} \ln a_{\alpha} (\partial \xi_{\alpha}) = 0,
\qquad   \alpha = 1,2.
\end{gather}

\begin{proposition} Suppose that the functions $\xi_{\alpha}$ have
unit modulus (3.8) and satisfy the overdetermined system
composed of equations (2.1) and (3.9). Then the reciprocal
functions $\xi_{\alpha}$ are solutions of equations (2.1) and (3.9).
\end{proposition}

\begin{proof} We wish to show that $\xi_{\alpha}$ is a solution of (3.9).
The derivatives of $\xi_{\alpha}^{-1}$ are given by
\begin{gather}
\partial (\xi_{\alpha}^{-1}) =- (\partial \xi_{\alpha}) \xi_{\alpha}^{-2},
\qquad
\bar{\partial} (\bar{\xi}^{-1}) =- (\bar{\partial} \bar{\xi}_{\alpha})
\xi_{\alpha}^{-2},\nonumber\\
\bar{\partial} \partial \xi_{\alpha}^{-1} = - \bar{\partial} \partial
\xi_{\alpha} \, \xi_{\alpha}^{-2} + 2 (\partial \xi_{\alpha})
(\bar{\partial} \xi_{\alpha}) \xi_{\alpha}^{-3}.
\end{gather}
Substituting (3.10) into (2.1) we obtain
\[
(-\partial \xi_{1}) \xi^{-2}_{1} (-\bar{\partial} \bar{\xi}_{1})
\bar{\xi}_{1}^{-2} = (-\partial \xi_{2}) \xi_{2}^{-2}
(- \bar{\partial} \bar{\xi}_{2}) \bar{\xi}_{2}^{-2}
\]
and, by straightforward computation, it is easy to show that, by using (3.8),
the above equation is satisfied whenever $\xi_{\alpha}$ satisfies (2.1).
Now we show that the reciprocal $\xi_{\alpha}^{-1}$ is a~solution of (3.9).
Substituting the derivatives (3.10) into (3.9) we obtain
\begin{gather}
-(\bar{\partial} \partial \xi_{\alpha}) \xi_{\alpha}^{-2} +
2 (\partial \xi_{\alpha})(\bar{\partial} \xi_{\alpha}) \xi_{\alpha}^{-3}
+ (- \partial \xi_{\alpha}) \xi_{\alpha}^{-2}
(- \bar{\partial} \bar{\xi}_{\alpha}) \bar{\xi}_{\alpha}^{-2}
\xi_{\alpha}^{-1} + 2 \bar{\partial} \ln a_{\alpha}
(- \partial \xi_{\alpha}) \xi_{\alpha}^{-2}
\nonumber\\
\qquad{}= \xi_{\alpha}^{-2}
[-(\bar{\partial} \partial \xi_{\alpha}) +2 (\partial \xi_{\alpha})
(\bar{\partial} \xi_{\alpha}) \xi_{\alpha}^{-1} + \xi_{\alpha}
(\partial \xi_{\alpha})( \bar{\partial} \bar{\xi}_{\alpha})
-2 \bar{\partial} \ln a_{\alpha} ( \partial \xi_{\alpha})].
\end{gather}
It is required to show that the expression (3.11) vanishes.
We assume that the $\xi_{\alpha}$ satis\-fy~(3.9). Then we can eliminate
the second derivative $\bar{\partial} \partial \xi_{\alpha}$
by using the second order equation~(3.9).
Moreover from (3.8) we have that $\xi_{\alpha} = \bar{\xi}_{\alpha}^{-1}$.
Differentiating both sides of this equation with respect to $\bar{\partial}$
we have
\[
\bar{\partial} \xi_{\alpha} = - (\bar{\partial} \bar{\xi}_{\alpha})
\bar{\xi}_{\alpha}^{-2}.
\]
Substituting the above equation for
$\bar{\partial} \xi_{\alpha}$ and (3.9) into
(3.11) we obtain
\begin{gather*}
(\partial \xi_{\alpha}) ( \bar{\partial} \bar{\xi}_{\alpha}) \xi_{\alpha}
+2 \bar{\partial} \ln a_{\alpha} (\partial \xi_{\alpha})
+2 (\partial \xi_{\alpha}) (\bar{\partial} \xi_{\alpha}) \xi_{\alpha}^{-1}
+ \xi_{\alpha} (\partial \xi_{\alpha})( \bar{\partial} \bar{\xi}_{\alpha})
-2 \bar{\partial} \ln a_{\alpha} (\partial \xi_{\alpha})
\\
\qquad{}= -2 (\partial \xi_{\alpha}) (\bar{\partial} \bar{\xi}_{\alpha}) \bar{\xi}_{\alpha}^{-1}
+2 \xi_{\alpha} (\partial \xi_{\alpha}) (\bar{\partial} \bar{\xi}_{\alpha})
=2 \bar{\xi}_{\alpha} (-(\partial \xi_{\alpha})(\bar{\partial} \bar{\xi}_{\alpha})
+  (\partial \xi_{\alpha})(\bar{\partial} \bar{\xi}_{\alpha}))=0,
\end{gather*}
which vanishes identically. This completes the proof.
\end{proof}

We now investigate the case in which all the derivatives of the
functions $\xi_{\alpha}$ are specified.

\begin{proposition} Let the functions $\xi_{\alpha}$
have unit modulus and their derivatives satisfy the following differential constraints
\begin{gather}
\partial \xi_{\alpha} = F_{\alpha} (z) \xi_{\alpha},
\qquad
\bar{\partial} \xi_{\alpha} =- \bar{F}_{\alpha} (\bar{z}) \xi_{\alpha},
\end{gather}
where the complex valued functions $F_{\alpha}(z)$ of class $C^{1}$
have equal modulus
\begin{gather}
|F_{1}(z)|^{2} = |F_{2} (z)|^{2}.
\end{gather}
Then the conditions (2.1) and (2.2) are satisfied identically and,
for any real constants $a_{\alpha}$,
the complex functions $\psi_{\alpha}$ and $\varphi_{\alpha}$ given by (2.14)
generate solutions of the KL system~(1.1).
\end{proposition}

\begin{proof} By substitution (3.12)
and their conjugates into (2.1) it is seen, using (3.8), that
equation (2.2) is reduced to (3.13). Differentiating (3.12) we have that
\begin{gather}
(i) \quad \bar{\partial} \partial \xi_{\alpha} = F_{\alpha} (z) \bar{\partial} \xi_{\alpha},
\qquad (ii)  \quad  \partial \bar{\partial} \xi_{\alpha} = - \bar{F}_{\alpha} (\bar{z})
\partial \xi_{\alpha}.
\end{gather}
Substituting the second derivative (3.14 i) and (3.12) into (2.16 i), we have
\[
\bar{\partial} \ln a_{\alpha} =
- \frac{F_{\alpha} (z) \bar{\partial} \xi_{\alpha}}{2 F_{\alpha}(z)
\xi_{\alpha}} - \frac{\bar{F}_{\alpha} (\bar{z}) \bar{\xi}_{\alpha}}{2}
\xi_{\alpha} = \frac{1}{2} \bar{F}_{\alpha} (\bar{z}) - \frac{1}{2}
\bar{F}_{\alpha} ( \bar{z}) = 0.
\]
By substitution the second derivative (3.14ii) and (3.12) in (2.16i),
it follows that
\[
\bar{\partial} \ln a_{\alpha} = - \frac{\bar{F}_{\alpha}(z) \partial \xi_{\alpha}}
{2 \partial \xi_{\alpha}} - \frac{\bar{F}_{\alpha} (z) \bar{\xi}_{\alpha}}{2} \xi_{\alpha} = 0.
\]
The compatibility condition (2.2) is satisfied identically
since $\bar{\partial} a_{\alpha} = 0$. So $a_{\alpha}$ is any real constant.
By virtue of Proposition 2 the complex functions $\psi_{\alpha}$
and $\varphi_{\alpha}$ which are defined in terms of $\xi_{\alpha}$
and $a_{\alpha}$ by (2.18) satisfy the KL system (1.1).
\end{proof}

We discuss now the possibility of constructing more general
classes of solutions of KL system (1.1) which are based on nonlinear
superpositions of elementary solutions of equation (3.9).

\begin{proposition} Consider two functionally independent
solutions, $\xi_{1 \alpha}$ and $\xi_{2 \alpha}$, of equations (3.9),
which are labelled with an additional index and $\alpha=1,2$.
Suppose that the complex functions
$\xi_{1 \alpha}$ and $\xi_{2 \alpha}$
have unit modulus
\[
|\xi_{\beta \alpha}|^{2} =1
\]
for $\alpha, \beta=1,2$. Suppose also that there exist real valued
functions, $a_{\alpha}(z, \bar{z})$, such that the equation
\begin{gather}
\bar{\partial} \partial \xi_{\beta \alpha}
+(\partial \xi_{\beta \alpha})(\bar{\partial}
\bar{\xi}_{\beta \alpha}) \xi_{\beta \alpha}
+ \bar{\partial} \ln a_{\alpha} (\partial \xi_{\beta \alpha})=0
\end{gather}
and its respective complex conjugate equation hold.
Then the products of the
functions
\begin{gather}
\eta_{\alpha}= \xi_{1 \alpha} \xi_{2 \alpha}
\end{gather}
have to satisfy the equations
\[
\partial \bar{\partial} \eta_{\alpha}
+ (\partial \eta_{\alpha}) (\bar{\partial} \bar{\eta}_{\alpha})
+ \bar{\partial} \ln a_{\alpha} (\partial \eta_{\alpha}) =0,
\qquad
\alpha=1,2.
\]
and their respective complex conjugate equations.
\end{proposition}

\begin{proof} We show that $\eta_{\alpha}$ given by (3.16)
satisfies (3.15). In fact differentiating the functions
$\eta_{\alpha}$ successively we obtain
\begin{gather*}
\partial \eta_{\alpha} =
(\partial \xi_{1 \alpha}) \xi_{2 \alpha} + \xi_{1 \alpha}
( \partial \xi_{2 \alpha}),
\qquad
\bar{\partial} \bar{\eta}_{\alpha} = (\bar{\partial}
\bar{\xi}_{1 \alpha}) \bar{\xi}_{2 \alpha}
+ \bar{\xi}_{1 \alpha} (\bar{\partial} \bar{\xi}_{2 \alpha}),
\\
\bar{\partial} \partial \eta_{\alpha}
= (\bar{\partial} \partial \xi_{1 \alpha}) \xi_{2 \alpha}
+(\partial \xi_{1 \alpha})(\bar{\partial} \xi_{2 \alpha})
+(\bar{\partial} \xi_{1 \alpha}) (\partial \xi_{2 \alpha})
+ \xi_{1 \alpha} (\bar{\partial} \partial \xi_{2 \alpha}).
\end{gather*}
Substituting the first and second derivatives of $\eta_{\alpha}$
into equation (2.15), we obtain
\begin{gather}
\xi_{2 \alpha} \bar{\partial} \partial \xi_{1 \alpha}
+ (\partial \xi_{1 \alpha})(\bar{\partial} \xi_{2 \alpha})
+ (\bar{\partial} \xi_{1 \alpha})
(\partial \xi_{2 \alpha}) + \xi_{1 \alpha}  \bar{\partial}
\partial \xi_{2 \alpha}+(( \partial \xi_{1 \alpha}) \xi_{2 \alpha} + \xi_{1 \alpha}
( \partial \xi_{2 \alpha}))
\nonumber\\
\qquad {}\times((\bar{\partial} \bar{\xi}_{1 \alpha})
\bar{\xi}_{2 \alpha} + \bar{\xi}_{1 \alpha}
( \bar{\partial} \bar{\xi}_{2 \alpha})) \xi_{1 \alpha} \xi_{2 \alpha}
+ \bar{\partial} \ln a_{\alpha}
(\partial \xi_{1 \alpha} \xi_{2 \alpha} + \xi_{1 \alpha}
\partial \xi_{2 \alpha})
\nonumber\\
\qquad {}= \xi_{2 \alpha}( \bar{\partial} \partial \xi_{1 \alpha}
+ \xi_{1 \alpha} (\partial \xi_{1 \alpha})(\bar{\partial}
\bar{\xi}_{ 1 \alpha}) +(\partial \xi_{1 \alpha}) \bar{\partial}
\ln a_{\alpha}) + \xi_{1 \alpha} (\bar{\partial} \partial
\xi_{2 \alpha} + \xi_{2 \alpha} (\partial \xi_{2 \alpha})
(\bar{\partial} \bar{\xi}_{2 \alpha}) \nonumber\\
\qquad {}+ \partial \xi_{2 \alpha}
\bar{\partial} \ln a_{\alpha})
+(\partial \xi_{1 \alpha})(\bar{\partial} \xi_{2 \alpha})
+ (\bar{\partial} \xi_{1 \alpha})(\partial \xi_{2 \alpha})
+ \xi_{1 \alpha}^{2} (\partial \xi_{2 \alpha})
(\bar{\partial} \bar{\xi}_{1 \alpha})\nonumber\\
\qquad{}+ \xi_{2 \alpha}^{2}(\partial \xi_{1 \alpha})(\bar{\partial}
\bar{\xi}_{2 \alpha}).
\end{gather}
Since the first two terms
appearing on the right hand side of (3.17)
are individually equation (3.15) with $\beta=1,2$ respectively,
these terms must vanish. So we show that the third term in (3.1)
also vanishes. This follows by differentiating the relation (3.8).
We obtain
\[
\bar{\partial} \bar{\xi}_{\beta \alpha} =- \xi_{\beta \alpha}^{-2}
(\bar{\partial} \xi_{\beta \alpha}).
\]
The third term on the right hand side of (3.17) vanishes
\[
(\partial \xi_{1 \alpha})(\bar{\partial} \xi_{2 \alpha})
+ (\bar{\partial} \xi_{1 \alpha})(\partial \xi_{2 \alpha})
-(\partial \xi_{2 \alpha})(\bar{\partial} \xi_{1 \alpha})
-(\partial \xi_{1 \alpha})(\bar{\partial} \xi_{2 \alpha})=0.
\]
This completes the proof.
\end{proof}

The fact that the functions $\xi_{\beta \alpha}$ individually
satisfy (2.1) for $\alpha, \beta =1,2$ is in itself not sufficient
to guarantee that the product functions $\eta_{\alpha} = \xi_{1
\alpha} \xi_{2 \alpha}$ will also satisfy equation~(2.1) due to
the presence of the first derivatives of the functions
$\eta_{\alpha}$.

\begin{proposition} Suppose that the functions $\xi_{j,\alpha}$, $j=1,2$,
satisfy condition (2.1) and have unit modulus (3.8).
Then the product functions
\begin{gather}
\eta_{\alpha} = \xi_{1 \alpha} \xi_{2 \alpha}
\end{gather}
for $\alpha=1,2$, satisfy condition
(2.1) provided that the following differential constraint
holds for the functions $\xi_{\beta \alpha}$
\begin{gather}
(\partial \xi_{21})(\bar{\partial} \bar{\xi}_{11}) \xi_{11} \bar{\xi}_{21}
+ \bar{\xi}_{11} \xi_{21} (\partial \xi_{11})(\bar{\partial}
\bar{\xi}_{21})\nonumber\\
\qquad {}= \xi_{12} \bar{\xi}_{22} (\partial \xi_{22})
(\bar{\partial} \bar{\xi}_{12}) + \xi_{22} \bar{\xi}_{12}
(\partial \xi_{12})(\bar{\partial} \bar{\xi}_{22}).
\end{gather}
\end{proposition}

\begin{proof} Suppose the set of functions $\xi_{11}$,
$\xi_{12}$ and $\xi_{21}$, $\xi_{22}$, satisfy (2.1). Then
\[
(\partial \xi_{11})(\bar{\partial} \bar{\xi}_{11})
= (\partial \xi_{12})(\bar{\partial} \bar{\xi}_{12}),
\qquad
(\partial \xi_{21})(\bar{\partial} \xi_{21})
= (\partial \xi_{22})(\bar{\partial} \bar{\xi}_{22}).
\]
Evaluating from (3.18) the expression $(\partial \eta_{1})
(\bar{\partial} \bar{\eta}_{1})-(\partial \eta_{2})
(\bar{\partial} \bar{\eta}_{2})$, which is
equivalent to (2.1) for the functions
$\eta_{\alpha}$, and next using equations (3.8), we obtain
\begin{gather*}
\partial (\xi_{11} \xi_{21}) \bar{\partial} (\bar{\xi}_{11}
\bar{\xi}_{21}) - \partial (\xi_{12} \xi_{22}) \bar{\partial}
(\bar{\xi}_{12} \bar{\xi}_{22})
=((\partial \xi_{11}) \xi_{21} + \xi_{11} (\partial \xi_{21}))\\
\qquad {}\times
(( \bar{\partial} \bar{\xi}_{11}) \bar{\xi}_{21}+ \bar{\xi}_{11}
(\bar{\partial} \bar{\xi}_{21}))
- ((\partial \xi_{12}) \xi_{22} + \xi_{12} (\partial \xi_{22}))
((\bar{\partial} \bar{\xi}_{12}) \bar{\xi}_{22}
+(\bar{\xi}_{12} \bar{\partial}) \bar{\xi}_{22})
\\
\qquad{}= (\partial \xi_{21})(\bar{\partial} \bar{\xi}_{11})
\xi_{11} \bar{\xi}_{21} + \bar{\xi}_{11} \xi_{21}
(\partial \xi_{11})(\bar{\partial} \bar{\xi}_{21})
- \xi_{12} \bar{\xi}_{22} (\partial \xi_{22})
(\bar{\partial} \bar{\xi}_{12})\\
\qquad{} - \xi_{22} \bar{\xi}_{12} (\partial \xi_{12})
(\bar{\partial} \bar{\xi}_{22}).
\end{gather*}
This vanishes whenever (3.19) is satisfied.
\end{proof}

Propositions 6 and 7 can be used to give a  criterion which
enable us to
determine whether a product of solutions of (3.9) is also a solution as well.
We denote the pair of functions $\xi_{i \alpha}$ in  the case
in which they are equal for $\alpha=1,2$ as follows
$\xi_{i1} = \xi_{i2} = \xi_{i}$.

Now we show that multisoliton solutions to the KL system (1.1) can
be constructed based on nonlinear superposition of $n$ elementary
solutions of the system composed of~(2.1) and (2.2).

\begin{proposition} (Factorization) Suppose that each function
$ \xi_{i}$ for $i=1, \ldots, n$
has unit modulus (3.8), and satisfies conditions
(2.1) and (2.2).
Suppose also that the functions $\xi_{i}$ are
solutions of the following differential constraint,
\begin{gather}
\bar{\partial} \partial \xi_{i \alpha} +2 \frac{(\partial \xi_{i \alpha})
(\bar{\partial} \bar{\xi}_{i \alpha})}{1 + |\xi_{i \alpha}|^{2}} \xi_{i \alpha}
+ 2 (\bar{\partial} \ln a ) (\partial \xi_{i \alpha}) = 0
\end{gather}
and its complex conjugate equation, where $a$ is a real-valued
function of $z$ and $\bar{z}$.
Then the product functions $\eta_{\alpha}$ defined by
\begin{gather}
\eta_{1} = \eta_{2} = \prod_{k=1}^{n} \xi_{k}
\end{gather}
satisfy the overdetermined system of equations (2.1), (2.2) and
(3.20). The functions $\eta_{\alpha}$ determine a multisolitonic
type solution to the KL system (1.1) by means of
equa\-tions~(2.18).
\end{proposition}

\begin{proof} The proof is by induction.
Consider two solutions $\xi_{1}$ and $\xi_{2}$,
each of unit modulus,
which both satisfy equation (3.20) under the same
real valued function $a (z, \bar{z})$ for both~$\xi_{i}$, $i=1,2$.
Then, from Proposition~6, the functions
$\eta_{\alpha}$, which are given by the product $\eta_{1}=\eta_{2}=
\xi_{1 \alpha} \xi_{2 \alpha}$, satisfy~(3.20)
under the same real valued function
$a(z, \bar{z})$. Moreover,
$\xi_{11}=\xi_{12}$ and $\xi_{21}=\xi_{22}$ holds.
The condition (2.1) is identically satisfied.
Equation (2.2) is the compatibility condition for the
function $a_{\alpha}$. Since the function $a( z, \bar{z})$ is
common for the entire set of $\xi_{\alpha}$, the compatibility
condition (2.2) is also satisfied identically.
Thus the product of $\xi_{1 \alpha}$ and $\xi_{2 \alpha}$
determines a solution to the KL system~(1.1) by Proposition~2.
By induction suppose that the functions
\[
q_{n-1, 1}= q_{n-1, 2} = \prod_{k=1}^{n-1}  \xi_{k, \alpha}
= \prod_{k=1}^{n} \xi_{k},
\]
satisfy the hypotheses of the theorem up to some $n$.
This means that $q_{n-1, \alpha} = q_{n-1}$ is a~solution of
(3.20) for the same real-valued function $a (z, \bar{z})$, and satisfies
(2.1) and (2.2). The new product function $\xi_{n, \alpha} = \xi_{n}$
also satisfies the hypotheses of Proposition~8, as well as
(2.1) and (2.2). Thus it follows by applying Proposition~6 that
the new product function
\[
\eta_{\alpha} = \left(\prod_{k=1}^{n-1} \xi_{k,\alpha}\right) \xi_{n,\alpha}
= \prod_{k=1}^{n} \xi_{k, \alpha}
\]
is also a solution of (3.19) under the same function $a(z, \bar{z})$.
This function $\eta_{\alpha}$ clearly has unit modulus. If we set
\[
\eta_{1} = \eta_{2} = \prod_{k=1}^{n} \xi_{k},
\]
then $\eta_{\alpha}$ satisfies (2.1), (2.2) as well as (3.10).
Hence the functions $\eta_{\alpha}$ can be used to ge\-ne\-ra\-te
new multi-soliton type solutions of KL system (1.1) by a
straightforward application of Proposition 2 to yield new
solutions in the form,
\begin{gather}
\varphi_{1}=\varphi_{2} = a(z, \bar{z})
\left(\partial \prod_{k=1}^{n-1} \xi_{k}\right)^{1/2}\!,
\quad
\psi_{1} = \psi_{2} = a (z, \bar{z}) \prod_{k=1}^{n-1}
\xi_{k} \left( \bar{\partial} \prod_{k=1}^{n-1} \bar{\xi}_{k}\right)^{1/2}\!.
\end{gather}
\end{proof}

\section{Reduction of the Konopelchenko--Landolfi system \\
to decoupled $\boldsymbol{CP^1}$ sigma models}

Now we discuss the case in which KL system (1.1) is subjected to a
single differential constraint and its conjugate. This allows us to reduce this system
to one which is composed of two decoupled $CP^1$ sigma model equations.

We start by introducing the new dependent variable
\begin{gather}
J = - \frac{1}{2} \left[ \frac{\partial \xi_{1} \partial \bar{\xi}_{1}}{(1 + |\xi_{1}|^{2})^{2}}
+ \frac{\partial \xi_{2} \partial \bar{\xi}_{2}}{(1 + |\xi_{2}|^{2})^{2}}\right].
\end{gather}

\begin{proposition} The overdetermined system composed of equations
(1.8) and the differential constraint $\bar{\partial} J = 0$ are consistent
if and only if the complex functions $\xi_{1}$ and $\xi_{2}$ satisfy the
decoupled pair of $CP^1$ sigma model equations
\begin{gather}
\bar{\partial} \partial \xi_{\alpha} - \frac{2 \bar{\xi}_{\alpha}}{1 + |\xi_{\alpha}|^{2}}
\bar{\partial} \xi_{\alpha} \partial \xi_{\alpha} = 0, \qquad   \alpha=1,2
\end{gather}
and their complex conjugates.
\end{proposition}

\begin{proof} Differentiation of $J$ with respect to $\bar{\partial}$
and imposing the differential constraint $\bar{\partial} J = 0$, we obtain the relation
\begin{gather}
\bar{\partial} J = - \frac{1}{2} \left[
\frac{\bar{\partial} \partial \xi_{1} \, \partial \bar{\xi}_{1}}
{(1 + |\xi_{1}|^{2})^{2}} + \frac{\partial \xi_{1} \bar{\partial} \partial \bar{\xi}_{1}}
{(1 + |\xi_{1}|^{2})^{2}} - 2 \frac{\partial \xi_{1} \partial \xi_{1}}{(1 + |\xi_{1}|^{2})^{3}}
( \bar{\partial} \xi_{1} \bar{\xi}_{1} + \xi_{1} \bar{\partial} \bar{\xi}_{1})\right.\nonumber\\
\left.\qquad{}+ \frac{\bar{\partial} \partial \xi_{2} \, \partial \bar{\xi}_{2}}
{(1 + |\xi_{2}|^{2})^{2}} + \frac{\partial \xi_{2} \,
\bar{\partial} \partial \bar{\xi}_{2}}{(1 + |\xi_{2}|^{2})^{2}}
-2 \frac{\partial \xi_{2} \partial \bar{\xi}_{2}}{(1 + |\xi_{2}|^{2})^{3}}
( \bar{\partial} \xi_{2} \bar{\xi}_{2} + \xi_{2}
\bar{\partial} \bar{\xi}_{2}) \right] = 0,
\end{gather}
and its complex conjugate. Solving the overdetermined system of equations
(1.8) and (4.3) for the second derivatives $\partial \bar{\partial} \xi_{1}$,
$\partial \bar{\partial} \xi_{2}$ and their conjugates, we find that
the functions $\xi_{1}$ and $\xi_{2}$ satisfy equations
(4.2) and their respective conjugates, respectively.
 \end{proof}

Note that all solutions of the $\mathbb C$$P^{1}$
model are well known~[16]. They fall into three
classes, those described by holomorphic or
antiholomorphic functions and the mixed ones.
The purpose for constructing solutions to the KL system~(1.1)
can be reduced to the following. Take any two
functionally independent solutions of the sigma model~(4.2)
 and substitute these solutions into equations
(2.18). Then the functions $a_{\alpha}$ are determined by
transformations (2.16). Hence,
by virtue of Propositions~2 and~9, the functions~$\psi_{\alpha}$ and~$\varphi_{\alpha}$ thus
obtained are solutions of the KL system~(1.1).

\section{Multisoliton solutions\\ of the Konopelchenko--Landolfi system}

At this point, we make use of the propositions
presented in Section~3 and~4 in order to construct
several new classes of solutions
to the KL system, including multisoliton solutions.

Now we discuss several classes of solutions
which can be obtained directly from
equations (2.1), (2.2) and transformation (2.18).

1. We start with a simple class of analytic solutions to equations
(2.1) and (2.2) of the exponential type
\[
\xi_{\alpha} = e^{z}.
\]
The successive derivatives of $\xi_{\alpha}$ are
\[
\partial \xi_{\alpha} = e^{z},
\qquad
\bar{\partial} \bar{\xi}_{\alpha} = e^{\bar{z}},
\qquad
\bar{\partial} \partial \xi_{\alpha} =0.
\]
Condition (2.1) is identically satisfied and,
from (2.16), the real functions $a_{\alpha}$
are determined by
\begin{gather}
\bar{\partial} \ln a_{\alpha} = -\frac{e^{\bar{z}}}
{1 + e^{z+\bar{z}}} e^{z} = - \bar{\partial}
\ln (1+ e^{z + \bar{z}}), \nonumber\\
\partial \ln a_{\alpha} =- \frac{e^{z}}{1+ e^{z + \bar{z}}} e^{\bar{z}}
=- \partial \ln (1 + e^{z + \bar{z}}).
\end{gather}
The compatibility condition for (5.1)
is satisfied identically. Thus equations (5.1)
are exact differentials and the functions $a_{\alpha}$
have the form
\[
a_{\alpha} ( z, \bar{z}) = c_{\alpha} (1 + e^{z +\bar{z}})^{-1},
\]
which are real-valued functions when $c_{\alpha}$ are real constants.

From Proposition 2 the functions
$\varphi_{\alpha}$ and $\psi_{\alpha}$ are given by
\begin{gather}
\varphi_{\alpha} = c_{\alpha} (1+ e^{z + \bar{z}}) e^{z/2},
\qquad
\psi_{\alpha} = c_{\alpha} (1+ e^{z + \bar{z}}) e^{z} e^{\bar{z}/2}.
\end{gather}
The equation for a CMC-surface immersed into ${\mathbb R}^{4}$
can be obtained by substituting
solutions (5.2) into (1.3).
By eliminating the parameter $t=e^{z+\bar{z}}$
from the pair of equations
\[
X_{1}^{2} + X_{2}^{2} = t (t-1)^{2} \left(3 + 3 t + t^{2}\right)^{2},
\qquad
X_{3}^{2} + X_{4}^{2} = 4 (1+ t)^{6}
\]
an algebraic equation is found which describes the surface
under consideration.

2. A more general class of exponential type solutions of (2.1) and (2.2)
are provided by the analytic functions $\xi_{\alpha}$ of the form
\begin{gather}
\xi_{\alpha} = e^{i \phi_{\alpha} (z, \bar{z})},
\end{gather}
where the $\phi_{\alpha}$ are real-valued functions.
The derivatives of $\xi_{\alpha}$ are given by
\[
\partial \xi_{\alpha} = i \partial \phi_{\alpha}
e^{i \phi_{\alpha}},
\qquad
\bar{\partial} \bar{\xi}_{\alpha} =- i
\bar{\partial} \phi_{\alpha} e^{-i \phi_{\alpha}}.
\]
By the use of (5.3), condition (2.1) becomes
\begin{gather}
(\partial \phi_{1})(\bar{\partial} \phi_{1})
= (\partial \phi_{2} )(\bar{\partial} \phi_{2})
\end{gather}
which holds whenever $\partial \phi_{1}$ and $\partial \phi_{2}$
differ only by a phase. Substituting (5.3) into equation~(2.16),
we get
\begin{gather}
\bar{\partial} \ln a_{\alpha} = - \frac{\partial \bar{\partial} \phi_{\alpha}}
{2 \partial \phi_{\alpha}} - \frac{i}{2} \bar{\partial} \phi_{\alpha}
+ \frac{i}{2} \bar{\partial} \phi_{\alpha} =- \frac{\partial
\bar{\partial} \phi_{\alpha}}{2 \partial \phi_{\alpha}}
\end{gather}
and its conjugate equation.
Thus the compatibility condition for (5.5) requires that
\begin{gather}
{\rm Im}\left( - \frac{\partial \partial \bar{\partial} \phi_{\alpha}}
{2 \partial \phi_{\alpha}} + \frac{\partial \bar{\partial} \phi_{\alpha}}
{2 (\partial \phi_{\alpha})^{2}} \, \partial^{2} \phi_{\alpha}\right)=0
\end{gather}
holds. Thus, the functions $\xi_{\alpha}$ will provide a solution
of (1.1) through (2.18) if condition~(5.5) as well as (5.1) holds.
In particular, if the function $\phi_{\alpha}$ is a real-valued
harmonic function,
\begin{gather}
\partial \bar{\partial} \phi_{\alpha} = 0,
\end{gather}
the compatibility condition (5.6) is satisfied identically.
Thus the corresponding solutions for the KL system (1.1) are
given by the following expressions
\begin{gather}
\varphi_{\alpha} = a_{\alpha} (i \partial \phi_{\alpha} e^{i
\phi_{\alpha}})^{1/2}, \qquad \psi_{\alpha} = a_{\alpha} e^{i
\phi_{\alpha}} ( -i \bar{\partial} \phi_{\alpha} e^{-i
\phi_{\alpha}})^{1/2}.
\end{gather}
For example, taking harmonic $\phi_{1}=\phi_{2}=(z^{2}+\bar{z}^{2})$,
the corresponding surface is a cylinder with $X_{3}$ as its symmetry axis.

3. Consider a monomial class of solutions
of (2.1) and (2.2) which are generated by $\xi_{\alpha}$
of the form
\begin{gather}
\xi_{\alpha} = \left[\frac{z - z_{0}}{\lambda_{\alpha}}\right]^{n},
\end{gather}
where $\lambda_{\alpha}$ and $z_{0}$ are arbitrary complex numbers.
The derivatives of (5.9) are given by
\begin{gather}
\partial \xi_{\alpha} = \frac{n}{\lambda_{\alpha}^{n}}
( z- z_{0})^{n-1},
\qquad
\bar{\partial} \bar{\xi}_{\alpha} = \frac{n}{\bar{\lambda}_{\alpha}^{n}}
(\bar{z} - \bar{z}_{0})^{n-1},   \qquad \bar{\partial} \partial \xi_{\alpha} = 0.
\end{gather}
It is easy to compute constraint (2.1), which will be satisfied
for all $z$ provided that $|\lambda_{\alpha}|^{2} = 1$ holds.
Substituting derivatives (5.10) into (2.16), we obtain
\begin{gather}
\bar{\partial} \ln a_{\alpha} = -  \frac{n \bar{\lambda}^{-n}_{\alpha}
(\bar{z}- \bar{z}_{0})^{n-1}}
{1 + |z - z_{0}|^{2n}}  \frac{(z- z_{0})^{n}}{\lambda_{\alpha}^{n}}
= - \bar{\partial} \ln \left(1 + |z - z_{0}|^{2n}\right),   \nonumber\\
       \partial \ln a_{\alpha} = -  \frac{n \lambda^{-n}_{\alpha}
(z- z_{0})^{n-1}}{1 + |z - z_{0}|^{2n}}
 \frac{(\bar{z} - \bar{z}_{0})^{n}}{\bar{\lambda}_{\alpha}^{n}}
= - \partial \ln \left(1 + |z - z_{0}|^{2n}\right).
\end{gather}
The compatibility condition for (5.11) is
identically satisfied and, integrating (5.11), we obtain
\begin{gather}
a_{\alpha} ( z, \bar{z}) =c_{\alpha} \left(1+ |z - z_{0}|^{2n}\right)^{-1},
\qquad
c_{\alpha} \in \mathbb R.
\end{gather}
Substituting (5.12) into (2.18), we obtain solutions $\varphi_{\alpha}$ and $\psi_{\alpha}$
of the KL system (1.1),
\begin{gather}
\varphi_{\alpha} = c_{\alpha} \left(1 + |z- z_{0}|^{2n}\right)^{-1}
\left(\frac{n}{\lambda_{\alpha}^{n}} (z - z_{0})^{n-1}\right)^{1/2},
\nonumber\\
\psi_{\alpha}=  c_{\alpha} \left(1 + |z - z_{0}|^{2n}\right)^{-1}
\left(\frac{z - z_{0}}{\lambda_{\alpha}}\right)^{n}
\left(\frac{n}{\bar{\lambda}_{\alpha}^{n}} (\bar{z} - \bar{z}_{0})^{n-1}\right)^{1/2}.
\end{gather}
By eliminating the parameter $t= |z- z_{0}|^{2n}$
from the following pair of equations
\[
X_{1}^{2} + X_{2}^{2} = (1+t)^{-2} t \left(1 - t^{-1}\right)^{2},
\qquad
X_{3}^{2} + X_{4}^{2} = 4 (1+t)^{-2}
\]
we can determine the associated surface for solutions (5.13).

4. Now we discuss the construction of a class of solution
of (2.1) and (2.2) which admits two arbitrary functions of one variable
$f_{\alpha} (z)$,
\begin{gather}
\xi_{\alpha} = \frac{f_{\alpha} (z)}{\bar{f}_{\alpha} (\bar{z})},
\qquad
\alpha=1,2.
\end{gather}
It is clear from (5.11) that $|\xi_{\alpha}|^{2}=1$ holds, and
the derivatives of (5.14) are given by
\[
\partial \xi_{\alpha} = \frac{\partial f_{\alpha}}{\bar{f}_{\alpha}},
\qquad
\bar{\partial} \partial \xi_{\alpha} =- \frac{\partial f_{\alpha}}
{\bar{f}_{\alpha}^{2}} \bar{\partial} \bar{f}_{\alpha}
\]
and their respective conjugates. Then, from equation (2.16), we find that
\begin{gather}
\bar{\partial} \ln a_{\alpha} = \frac{1}{2} \frac{\partial f_{\alpha}}
{\bar{f}_{\alpha}^{2}} \bar{\partial} \bar{f}_{\alpha}
\frac{\bar{f}_{\alpha}}{\partial f_{\alpha}}
- \frac{1}{2} \frac{\bar{\partial} \bar{f}_{\alpha}}{f_{\alpha}}
\frac{f_{\alpha}}{\bar{f}_{\alpha}}
= \frac{1}{2} \left( \frac{\bar{\partial} \bar{f}_{\alpha}}{\bar{f}_{\alpha}}
- \frac{\bar{\partial} \bar{f}_{\alpha}}{\bar{f}_{\alpha}}\right)=0,
\end{gather}
as well as its conjugate. So $a_{\alpha}$ is any real constant.
Thus the compatibility condition for~(5.15) is satisfied
identically and  equation (2.1) is satisfied provided that
\[
\frac{|\partial f_{1}|^{2}}{|f_{1}|^{2}}
= \frac{|\partial f_{2}|^{2}}{|f_{2}|^{2}}
\]
holds.
Thus, from (2.18), for any real constants $a_{\alpha}$ we have
\[
\varphi_{\alpha} = a_{\alpha} \left(\frac{\partial f_{\alpha}}{\bar{f}_{\alpha}
}\right)^{1/2},
\qquad
\psi_{\alpha} = a_{\alpha} \frac{f_{\alpha}}{\bar{f}_{\alpha}}
\left(\frac{\bar{\partial} \bar{f}_{\alpha}}{f_{\alpha}}\right)^{1/2}.
\]
As an example, if we
take $f(z)=z^{n}$, from (1.3) the corresponding surface can be calculated
explicitly and is a cylinder with symmetry axis $X_{3}$.

5. Consider a class of rational
solutions of the KL system (1.1) which admits simple poles
based on transformation (1.5),
\begin{gather}
\xi_{\alpha} = \frac{z - b_{\alpha}}{\bar{z} - \bar{b}_{\alpha}},
\qquad
\alpha=1,2, \qquad b_{\alpha} \in \mathbb C.
\end{gather}
Equations (5.16) obey the constraint $|\xi_{\alpha}|^{2} =1$.
The derivatives of $\xi_{\alpha}$ are given by
\begin{gather}
\partial \xi_{\alpha} = \frac{1}{\bar{z} - \bar{b}_{\alpha}},
\qquad
\bar{\partial} \bar{\xi}_{\alpha} = \frac{1}{z - b_{\alpha}},
\qquad
\bar{\partial} \partial \xi_{\alpha} =
- \frac{1}{(\bar{z} - \bar{b}_{\alpha})^{2}}.
\end{gather}
Then condition (2.1) is reduced to the following
\begin{gather}
(z - b_{1})(\bar{z} - \bar{b}_{1}) = (z - b_{2})(\bar{z} - \bar{b}_{2}).
\end{gather}
Equation (5.18) is satisfied when $b_{1} = b_{2} =b$.
Equation (2.16) reduces to
\begin{gather}
\partial a_{\alpha}=0, \qquad
\bar{\partial} a_{\alpha} =0
\end{gather}
and $a_{\alpha}$ is any real constant.
Hence, applying Proposition~2, we obtain solutions of the KL system
(1.1) of the form
\begin{gather}
\varphi_{\alpha}= a_{\alpha}
\left(\frac{1}{\bar{z} - \bar{b}}\right)^{1/2},
\qquad
\psi_{\alpha} = a_{\alpha} \left(\frac{z - b}{\bar{z} - \bar{b}}\right)
\left(\frac{1}{z - b}\right)^{1/2}.
\end{gather}
The surface in this case is a cylinder with $X_{3}$ the symmetry axis.

6. By the use of Proposition 2, an interesting class of periodic solutions
to the KL system (1.1) satisfying the algebraic constraint (3.8)
can be constructed. This class of solution can be determined by
periodic functions $\xi_{\alpha}$ of the form
\begin{gather}
\xi_{\alpha} = \exp( \cos(z-b_{\alpha}) - \cos(\bar{z} - \bar{b}_{\alpha})),
\qquad
b_{\alpha} \in \mathbb C, \qquad \alpha=1,2,
\end{gather}
which satisfy (3.8). Their successive derivatives are given by
\begin{gather}
\partial \xi_{\alpha} =- \sin(z-b_{\alpha}) \exp
(\cos (z-b_{\alpha}) - \cos( \bar{z} - \bar{b}_{\alpha})),
\nonumber\\
\bar{\partial} \bar{\xi}_{\alpha} = - \sin (\bar{z} - \bar{b}_{\alpha})
\exp( \cos(\bar{z} - \bar{b}_{\alpha})-\cos(z- b_{\alpha}),
\nonumber\\
\bar{\partial} \partial \xi_{\alpha} = \sin(z- b_{\alpha})
\sin(\bar{z} - \bar{b}_{\alpha})
\exp(\cos( z - b_{\alpha} )- \cos(\bar{z}- \bar{b}_{\alpha})).
\end{gather}
Equation (2.1) becomes
\begin{gather}
|\sin( z - b_{1})|^{2} =
|\sin( z - b_{2})|^{2}.
\end{gather}
Equation (5.23) holds when $b_{1}=b_{2}=b$.
Substituting the derivatives (5.22) into (2.16), we find that
the functions $a_{\alpha}$ satisfy the following conditions
\begin{gather}
\partial \ln a_{\alpha} = \sin(\bar{z}- \bar{b}_{\alpha}),
\qquad
\bar{\partial} \ln a_{\alpha} = \sin (z - b_{\alpha}).
\end{gather}
Integrating (5.24), we get
\begin{gather}
a_{\alpha} = \exp( c_{\alpha} (\sin(z- b_{\alpha})
+ \sin( \bar{z}- \bar{b}_{\alpha}))).
\end{gather}
If $b_{1}=b_{2}=b \in \mathbb C$, equations (2.18)
lead to the following nontrivial, periodic solutions of the KL system (1.1),
\begin{gather}
\varphi_{\alpha} = \exp(c_{\alpha} (\sin(z-b)+\sin(\bar{z}- \bar{b})))
(-\sin(z-b) \exp(\cos(z-b)-\cos(\bar{z} - \bar{b})))^{1/2},
\nonumber\\
\psi_{\alpha} = \exp(c_{\alpha}(\sin(z-b)+\sin(\bar{z}- \bar{b})))
\exp(\cos(z-b)- \cos(\bar{z} - \bar{b}))\nonumber\\
\phantom{\psi_{\alpha} =}{}\times
(-\sin(\bar{z} - \bar{b}) \exp(\cos(\bar{z} -\bar{b})
-\cos(z-b)))^{1/2}.
\end{gather}
The corresponding surface is a cylinder with $X_{3}$ as symmetry axis.

7. Another class of solutions of the KL system can
be obtained by replacing the cosine function in (5.21) by
the hyperbolic function $\cosh$ such that functions $\xi_{\alpha}$
satisfy (3.8)
\begin{gather}
\xi_{\alpha} = \exp(\cosh(z - b_{\alpha}) - \cosh( \bar{z}
- \bar{b}_{\alpha})).
\end{gather}
The successive derivatives of $\xi_{\alpha}$ are given by
\begin{gather}
\partial \xi_{\alpha} = \sinh ( z - b_{\alpha})
\exp( \cosh ( z- b_{\alpha}) - \cosh( \bar{z} - \bar{b}_{\alpha})),
\nonumber\\
\bar{\partial} \bar{\xi}_{\alpha} = \sinh (\bar{z} - \bar{b}_{\alpha})
\exp ( \cosh (\bar{z} - \bar{b}_{\alpha}) - \cosh ( z - b_{\alpha})),
\nonumber\\
\bar{\partial} \partial \xi_{\alpha} = - \sinh( \bar{z} - \bar{b}_{\alpha})
\sinh ( z - b_{\alpha}) \exp( \cosh ( z - b_{\alpha}) -
\cosh ( \bar{z} - \bar{b}_{\alpha})).
\end{gather}
Condition (2.1) then takes the form
\begin{gather}
|\sinh ( z - b_{1}) \exp( \cosh (z - b_{1}) -
\cosh ( \bar{z} - \bar{b}_{1}))|\nonumber\\
\qquad {}=
|\sinh ( z - b_{2}) \exp( \cosh ( z - b_{2}) - \cosh
(\bar{z} - \bar{b}_{2})) |.
\end{gather}
Equation (5.29) is satisfied when $b_{1} = b_{2} = b \in \mathbb C$.
Equations (2.16) become identical to (5.19) in this case.
Hence $a_{\alpha}$ are any real constants.
Therefore, by the use of (2.18) the
required solutions of KL
system (1.1) are given by
\begin{gather}
\varphi_{1,2} = a_{\alpha}
( \sinh (z- b) \exp (\cosh (z - b) - \cosh (\bar{z} - \bar{b})))^{1/2},
\nonumber\\
\psi_{1,2} = a_{\alpha} \exp( \cosh (z - b) - \cosh (\bar{z} - \bar{b}))
\nonumber\\
\phantom{\psi_{1,2}=}{}\times(\sinh (\bar{z} - \bar{b}) \exp( \cosh ( \bar{z} - \bar{b})
- \cosh ( z - b)))^{1/2}.
\end{gather}
These solutions represent a bump type solution and the
corresponding surface is a cylinder with symmetry axis $X_{3}$.
Note that solutions which yield cylinders, such as (5.8), (5.20),
(5.26) and (5.30), have applications to certain types
of cosmological models and are useful for describing
event horizons in general relativity~[4].

8. Another class of hyperbolic solutions is obtained
by using the $\tanh$ function instead of $\cosh$ in expression (5.21)
and is generated by
\[
\xi_{\alpha} = \exp( \tanh ( z- b_{\alpha}) -
\tanh ( \bar{z} - b_{\bar{\alpha}})).
\]
The successive derivatives of $\xi_{\alpha}$ are given by
\begin{gather*}
\partial \xi_{\alpha} = \mbox{sech}^{2} ( z- b_{\alpha})
\exp (\tanh ( z - b_{\alpha}) - \tanh (\bar{z} - \bar{b}_{\alpha})),
\\
\bar{\partial} \bar{\xi}_{\alpha} =
\mbox{sech}^{2} ( \bar{z} - \bar{b}_{\alpha})
\exp ( \tanh (\bar{z} - \bar{b}_{\alpha})
- \tanh ( z - b_{\alpha})),
\\
\bar{\partial} \partial \xi_{\alpha} =
- \mbox{sech}^{2} ( z - b_{\alpha}) \mbox{sech}^{2} (\bar{z} - \bar{b}_{\alpha})
\exp( \tanh (z - b_{\alpha}) - \tanh (\bar{z} - \bar{b}_{\alpha})).
\end{gather*}
Condition (2.1) takes the form
\begin{gather}
|{\rm sech}^{2} ( z - b_{1}) \exp (\tanh (z - b_{1})
- \tanh (\bar{z} - \bar{b}_{1}))|\nonumber\\
\qquad {}= |{\rm sech}^{2} ( z - b_{2}) \exp( \tanh (z - b_{2})
- \tanh ( \bar{z} - \bar{b}_{2}))|
\end{gather}
which holds whenever $b_{1}=b_{2}=b$.
Equations (2.16) are reduced to (5.19),
which implies that the $a_{\alpha}$ are
real constants. Thus from (2.18) we can write
\begin{gather*}
\varphi_{\alpha} = a ({\rm sech}^{2} (z -b)
\exp (\tanh ( z - b) - \tanh (\bar{z} - \bar{b})))^{1/2}, \nonumber\\
\psi_{\alpha} = a \exp( \tanh (z - b) - \tanh (\bar{z} - \bar{b})))\nonumber\\
\phantom{\psi_\alpha=}{}\times
({\rm sech}^{2} ( \bar{z} - \bar{b}) \exp
( \tanh (\bar{z} - \bar{b}) - \tanh ( z - b)))^{1/2},\qquad
\alpha=1,2.
\end{gather*}
This type of solution represents a kink-type solution.
Substituting these in (1.3), the corresponding surface is given
by a cylinder with $X_{3}$ as its
symmetry axis.

9. Finally we discuss a class of solutions
for the KL system (1.1) admitting simple poles which can be
constructed by applying Proposition 8, namely,
\begin{gather}
\xi_{k1} = \xi_{k2} = \frac{z - b_{k}}{\bar{z} - \bar{b}_{k}},
\qquad b_{k} \in \mathbb C.
\end{gather}
These functions $\xi_{k \alpha}$ have unit
modulus and identically satisfy condition (2.1) and
equation~(3.20) for any real constants $a_{\alpha}$.
By virtue of Proposition 8, a solution of KL system~(1.1) is generated by
the functions $\eta_{\alpha}$ in the form of the product,
\begin{gather}
\eta_{1} = \eta_{2} = \prod_{k=1}^{n} \frac{z - b_{k}}
{\bar{z} - \bar{b}_{k}}.
\end{gather}
Substituting (5.33) into (2.18), we obtain
\[
\varphi_{\alpha} = a \left(\eta_{\alpha}
\sum_{k=1}^{n} \frac{1}{z - b_{k}}\right)^{1/2},
\qquad
\psi_{\alpha} = a \eta_{\alpha}
\left( \bar{\eta}_{\alpha} \sum_{k=1}^{n} \frac{1}{\bar{z} - \bar{a}_{k}}\right)^{1/2}.
\]
Note that this type of solution admits only simple poles and
represents multisoliton solutions of the KL system (1.1).
The associated surface for $n=1$ is a cylinder.

With regard to the examples presented here, we mention
two configurations which are different from instantons and have
applications in string theory~[4].

The developable surfaces are the simplest. These surfaces have
Gaussian curvature $K=0$, satisfy (1.1) and the additional
constraint
\begin{gather}
\left(|\psi_{1}|^2 + |\varphi_{1}|^2 \right)
\left(|\psi_{2}|^2 + |\varphi_{2}|^2 \right) = |A(z)|^2,
\end{gather}
where $A(z)$ is an arbitrary holomorphic function.

A second class of surfaces, those with flat normal bundle,
correspond to vanishing normal curvature $K_{N}=0$. These
surfaces are generated by the system (1.1) subject to the
additional constraint
\begin{gather}
\left(|\psi_{2}|^2 + |\varphi_{2}|^2\right)
= |A(z)|^{2} \left( |\psi_{1}|^{2} + | \varphi_{1}|^2\right),
\end{gather}
where $A(z)$ is an arbitrary holomorphic function.
If $A(z)=1$ and we differentiate both sides of (5.35)
with respect to $\partial$ and use (1.1) to eliminate
the known derivatives of $\psi_{i}$ and $\varphi_{i}$,
we obtain the constraint
\[
\bar{\partial} \psi_{1} \bar{\psi}_{1} + \varphi_{1}
\bar{\partial} \bar{\varphi}_{1}
= \bar{\partial} \psi_{2} \bar{\psi}_{2} + \varphi_{2}
\bar{\partial} \varphi_{2}.
\]
Several solutions which have been presented in this Section
fall into this second classification if the functions
$a_{\alpha}$ are chosen properly.
For example, the solutions (5.20) and (5.30) in
examples~5 and~7 satisfy (5.35) for $A(z)=1$
provided that $a_{1}=a_{2}=1$. Another class of
solution (5.26) obeys condition (5.35) when
$c_{1} = c_{2} =1$.

\section{A simple model for strings in $\boldsymbol{{\mathbb R}^4}$}

According to~[4], the unification of gravity and quantum mechanics
seems to require a~new formulation of physics at small distance
scales. In the string approach elementary particles can be thought
of as strings, and differ from all familiar quantum mechanical
field theories the constituent particles of which are pointlike,
whereas a string has extension in space-time [21, 22]. Moreover,
superstring theory combines string theory with another
mathematical structure namely supersymmetry. This theory makes it
possible to consider all four fundamental forces as various
aspects of a single underlying principle~[23]. The forces are
unified in a way determined by the requirement that the theory be
internally consistent. The strings which are postulated by the
theory would then be about $10^{20}$ times smaller than the
diameter of the proton.

We introduce an independent world sheet metric,
$g_{ab} (x,y)$, which is
independent of the string variables. The Polyakov form
of the Lagrangian density~[24] is written as
\begin{gather}
L = -  \frac{1}{4 \pi \alpha}
\sqrt{-g} g^{ab} \partial_{a} X_{\mu} \partial_{b} X^{\mu}, \qquad
a,b = 1,2  \qquad \mu=1, \ldots, 4,
\end{gather}
with action
\[
S = \int dx \, dy \, L,
\]
where $g= \det(g_{ab})$, and $\alpha$ is the square of the
characteristic length scale of perturbative string theory.
The tension of the fundamental string is $1/ 2 \pi \alpha$.
This is a generalization of the second order point-particle
action. Notice that the Polyakov action resembles an action
with scalar fields interacting with an external two-dimensional
gravitational field~[24]. It is worth noting here that the
Polyakov form of the action is in fact equivalent
to another form of the action which appears in string theory~[24],
namely, the Nambu--Goto action. To obtain this equivalence,
the equation of motion which is derived by varying the
metric can be used. This implies that
\begin{gather}
h_{ab}  (-h)^{-1/2} = g_{ab} (-g)^{-1/2},
\end{gather}
where $h_{ab}$ is the induced metric, so that $g_{ab}$
is proportional to the induced metric. Equation~(6.2)
can be used to eliminate $g_{ab}$ from the Polyakov
action, thus giving the Nambu--Goto form.
The independent world sheet metric $g_{ab}$ is taken to be
\[
g_{ab} = g^{ab} = \left(  \begin{array}{cc}
0  &   1   \\
1  &   0    \\
\end{array}   \right).
\]
The Euler Lagrange equations of motion are obtained from (6.1)
and have the form,
\begin{gather}
\partial_{x} \partial_{y} X_{\mu} = 0, \qquad \mu = 1, \ldots, 4.
\end{gather}
Applying the identity $\partial_{x} \partial_{y} = i (\partial^{2} -
\bar{\partial}^{2})$, equation (6.2) becomes
\begin{gather}
(\partial^{2} - \bar{\partial}^{2}) X_{\mu} = 0.
\end{gather}
Substituting the expressions for the position vector ${\bf X}$
given by (1.3) into (6.4), we obtain
the following set of equations
\begin{gather}
\partial (\bar{\psi}_{1} \bar{\psi}_{2}
+ \varphi_{1} \varphi_{2}) + \bar{\partial}
( \psi_{1} \psi_{2} + \bar{\varphi}_{1} \bar{\varphi}_{2}) = 0,
\nonumber\\
\partial ( \bar{\psi}_{1} \bar{\psi}_{2} -
\varphi_{1} \varphi_{2}) - \bar{\partial}
( \psi_{1} \psi_{2} - \bar{\varphi}_{1}
\bar{\varphi}_{2}) = 0,
\nonumber\\
\partial ( \bar{\psi}_{1} \varphi_{2} +
\bar{\psi}_{2} \varphi_{1}) - \bar{\partial}
(\psi_{1} \bar{\varphi}_{2} + \psi_{2}
\bar{\varphi}_{1} ) = 0,
\nonumber\\
\partial ( \bar{\psi}_{1} \varphi_{2} -
\bar{\psi}_{2} \varphi_{1}) + \bar{\partial}
(\psi_{1} \bar{\varphi}_{2} - \psi_{2}
\bar{\varphi}_{1}) = 0.
\end{gather}
Adding the first two and last two equations in (6.5),
respectively, we obtain
\begin{gather}
\partial (\bar{\psi}_{1} \bar{\psi}_{2})
+ \bar{\partial} ( \bar{\varphi}_{1} \bar{\varphi}_{2} ) = 0,
\qquad
\partial ( \bar{\psi}_{1} \varphi_{2})
- \bar{\partial} ( \psi_{2} \bar{\varphi}_{1} ) = 0.
\end{gather}
Equations (6.6), together with their complex conjugates,
are equivalent to (6.5). For example, taking the sum and difference
of the first equation in (6.6) with its own conjugate, we obtain the first two
equations in (6.5).

Now in this context we discuss a condition under which
the KL system (1.1) becomes a linear system of equations.

\begin{proposition} Consider the overdetermined system composed of
the KL-system~(1.1)
which is subjected to DCs of the form
\begin{gather}
\psi_{1} \partial \bar{\psi}_{1} - \bar{\varphi}_{1} \partial \varphi_{1}
+ \psi_{2} \partial \bar{\psi}_{2} - \bar{\varphi}_{2} \partial
\varphi_{2} = 0.
\end{gather}
Assume that all first order derivatives of $\psi_{\alpha}$ and
$\varphi_{\alpha}$ with respect to $z$ and $\bar{z}$
are expressible in terms of polynomial functions
which depend on $\psi_{\alpha}$ and $\varphi_{\alpha}$ with
constant coefficients.
Then the KL system (1.1) subjected to (6.7) is equivalent to the
following linear system of equations
\begin{gather}
\partial \left(   \begin{array}{c}
\psi_{\alpha}    \\
\varphi_{\alpha} \\
\end{array}      \right)
= p_{0} \left(   \begin{array}{c}
\varphi_{\alpha}  \\
- \epsilon \psi_{\alpha}  \\
\end{array} \right),  \qquad
\bar{\partial}  \left(   \begin{array}{c}
\psi_{\alpha}    \\
\varphi_{\alpha}  \\
\end{array}       \right)
= p_{0} \left(    \begin{array}{c}
\epsilon \varphi_{\alpha}  \\
- \psi_{\alpha}   \\
\end{array}       \right),   \qquad \epsilon = \pm 1,
\\
(u_{1} u_{2})^{1/2} = p_{0},   \qquad
|\psi_{\alpha}|^{2} + |\varphi_{\alpha}|^{2} = u_{\alpha},
\end{gather}
where the quantities $p_{0}$ and $u_{\alpha}$ are
arbitrary real constants.
\end{proposition}

\begin{proof} {\samepage The aim is to find the explicit form of all first
derivatives of $\psi_{\alpha}$ and $\varphi_{\alpha}$ in such
a~way that they do not provide any additional differential
constraints on $\psi_{\alpha}$ and $\varphi_{\alpha}$ other than
those in (6.7) when the compatibility conditions are added. In
this case we can close the KL system (1.1) and show that the
compatibility conditions for (1.1) and (6.7) under the assumption
of Proposition~9 coincide only with the requirements (6.8) and
(6.9). Indeed, using $u_{\alpha}$ defined in (1.1) the first and
second derivatives of $u_{\alpha}$, $\psi_{\alpha}$ and
$\varphi_{\alpha}$ are
\begin{gather}
\partial u_{\alpha} = \psi_{\alpha} \partial \bar{\psi}_{\alpha}
+ \bar{\varphi}_{\alpha} \partial \varphi_{\alpha},  \nonumber\\
\partial \bar{\partial} u_{\alpha} = \partial \bar{\psi}_{\alpha}
\bar{\partial} \psi_{\alpha} + \partial \varphi_{\alpha}
\bar{\partial} \bar{\varphi}_{\alpha} - p^{2} u_{\alpha}, \qquad
\alpha=1,2,
\end{gather}
and
\begin{gather}
\bar{\partial} \partial \psi_{\alpha} = \bar{\varphi}_{\alpha}
\partial p - p^{2} \bar{\psi}_{\alpha}, \qquad
\partial^{2} \psi_{\alpha} = \varphi_{\alpha} \partial p + p
\partial \varphi_{\alpha},\nonumber\\
\bar{\partial} \partial \varphi_{\alpha} = - \psi_{\alpha}
\partial p - p^{2} \varphi_{\alpha}, \qquad  \bar{\partial}^{2}
\varphi_{\alpha} = - \psi_{\alpha} \bar{\partial} p - p
\bar{\partial} \psi_{\alpha}
\end{gather}
and their respective complex conjugate equations.}

Suppose that the unknown derivatives $(\bar{\partial} \psi_{\alpha},
\partial \bar{\psi}_{\alpha}, \partial \varphi_{\alpha},
\bar{\partial} \bar{\varphi}_{\alpha})$ other than those
appearing in (1.1) are polynomial in $\psi_{\alpha}$ and
$\varphi_{\alpha}$ with constant coefficients. The analysis
of the dominant terms in the variables $\psi_{\alpha}$
and $\varphi_{\alpha}$ for the compatibility conditions of
all first order derivatives leads to the requirement that all
unknown derivatives have to be cubic in terms of the fields
$\psi_{\alpha}$ and $\varphi_{\alpha}$ and take the following
specific form
\begin{gather}
\bar{\partial} \psi_{\alpha} =p \sum_{\alpha=1}^{2}
( c_{\alpha}^{1} \psi_{\alpha} + c_{\alpha}^{2} \varphi_{\alpha}),
\qquad
\partial \varphi_{\alpha} = p \sum_{\alpha=1}^{2}
( c_{\alpha}^{3} \psi_{\alpha} + c_{\alpha}^{4} \varphi_{\alpha}).
\end{gather}
Here, $c_{\alpha}^{i}$, $i = 1, \ldots, 4$, are complex constants
to be determined from the compatibility conditions for (6.11) and
(6.12). This leads us to a system of equations which are linear
in~$\psi_{\alpha}$ and $\varphi_{\alpha}$ and has a unique
solution of the form (6.7). Differentiation of $p$ with respect to
$z$ and $\bar{z}$ yields
\begin{gather}
\partial p = \frac{1}{2} p^{-1} (u_{2} \partial u_{1} + u_{1}
\partial u_{2}),  \qquad
\bar{\partial} p = \frac{1}{2} p^{-1} (u_{2} \bar{\partial} u_{1}
+ u_{1} \bar{\partial} u_{2}).
\end{gather}
Taking into account (6.8), we obtain that equations (6.13)
vanish identically and conditions~(6.9) hold. Note that under
the condition (6.7), we show that the KL system (1.1) admits
a conserved quantity with $p$ a real constant. It implies by
virtue of (1.4) that the Gaussian curvature $K=0$ and so
the surface is flat.
\end{proof}

Note that on substitution of the derivatives of functions
$\psi_{\alpha}$ and $\varphi_{\alpha}$, given in (6.8) into the
string equations (6.6), it is seen that (6.8) is a solution to the
string equations. If time is regarded as a spatial dimension, as
appropriate in $\mathbb R^4$, the world sheet of a closed string
can be thought of as a surface that joins the string at its
initial point and at the end of its spacetime path. It has been
shown here that the Polyakov form of the action~[4,~25] can be
used in the KL formulation of surfaces in $\mathbb R^{4}$, where
the motion of the string is such that the action is minimized.
This variational principle yields the string equations of
motion~(6.6).

In conclusion it is worth noting that the study of
stable classical solutions presented here is one of the
most important problems in investigating quantum theories
of strings, in particular, in constructing perturbation series
about a known classical solution, $X_{\mu} \rightarrow
X_{\mu, cl} + X_{\mu,q}$.

From the physical point of view, to perform calculations
in the path integral approach, one often shifts the
integration variable by a solution to the classical system.
In order to extract physical predictions from a theory
we must firstly quantize the theory and obtain the physical
states. On account of symmetries present in such theories,
there can be great redundancies in the degrees of
freedom and so the quantization of the string can be
nontrivial.
A quantum theory which describes
interactions of strings can be represented by a Feynman
path integral. In this form of quantization the
amplitudes are obtained by summing over all possible
histories which interpolate between the initial and
final states. Each path is weighted by the factor
$\exp( i S_{cl})$, where $S_{cl}$ is the classical
action for the given history. An amplitude in string theory
is defined by summing over all world-sheets connecting
the initial and final curves, or surfaces. In the sum over
world sheets the integral runs over all Euclidean
metrics and over all embeddings of the surface
determined by the position vector $X^{\mu}$ of the
world sheet in spacetime.
It is usual to take the action in Euclidean form,
which is more accurately defined as far as convergence
of the path integral is concerned.

An analysis of solutions carried out in Section 4 can provide us
with admissible surfaces with respect to the KL system (1.1)
in $\mathbb R^4$, which can be used in the
calculation of quantum corrections to classical results.
Recently the GW representations for inducing minimal
surfaces in pseudo-Riemannian multidimensional spaces
has been formulated in~[1,~15].
It should be possible to extend the path integral
approach to these multidimensional spaces as well.
This task will be undertaken in future work.

\subsection*{Acknowledgments}

The authors thank Professor B~Konopelchenko
(Univestia di Lecce) for helpful and interesting discussions
on the topic of this paper.
This work was supported by a research grant from NSERC of
Canada and Fonds FCAR du Gouvernment du Qu\'ebec.

\label{bracken-lastpage}

\end{document}